\documentclass[twocolumn]{aastex631}
\usepackage{amsmath}

\begin{document}

\title{Tracing Structure: Shape and Centroid Deviations in 39 Strong Lensing Clusters as a Test of Cluster Formation Predictions}

\author[0009-0004-7337-7674]{Raven Gassis}
\affiliation{Department of Physics, University of Cincinnati, Cincinnati, OH 45221, USA}

\author[0000-0003-1074-4807]{Matthew B. Bayliss}
\affiliation{Department of Physics, University of Cincinnati, Cincinnati, OH 45221, USA}

\author[0000-0002-7559-0864]{Keren Sharon}
\affiliation{Department of Astronomy, University of Michigan, 1085 South University Avenue, Ann Arbor, MI 48109, USA}

\author[0000-0003-3266-2001]{Guillaume Mahler}
\affiliation{STAR Institute, Quartier Agora -- All\'ee du six Ao\^ut 19c, B-4000 Li\`ege, Belgium}

\author[0000-0003-1370-5010]{Michael D. Gladders}
\affiliation{Department of Astronomy and Astrophysics, University of Chicago, 5640 South Ellis Avenue, Chicago, IL 60637, USA}
\affiliation{Kavli Institute for Cosmological Physics, University of Chicago, 5640 South Ellis Avenue, Chicago, IL 60637, USA}

\author[0000-0001-5226-8349]{Michael McDonald}
\affiliation{Department of Physics, Massachusetts Institute of Technology, Cambridge, MA 02139, USA}
\affiliation{Kavli Institute for Astrophysics and Space Research, Massachusetts Institute of Technology, 77 Massachusetts Avenue, Cambridge, MA 02139, USA}

\author[0000-0003-2200-5606]{H\aa kon Dahle}
\affiliation{Institute of Theoretical Astrophysics, University of Oslo, P.O. Box 1029, Blindern, NO-0315 Oslo, Norway}

\author[0000-0001-5097-6755]{Michael K. Florian}
\affiliation{Steward Observatory, University of Arizona, 933 North Cherry Ave., Tucson, AZ 85721, USA}

\author[0000-0002-7627-6551]{Jane R. Rigby}
\affiliation{Observational Cosmology Lab, Code 665, NASA Goddard Space Flight Center, Greenbelt, MD 20771, USA}

\author[0009-0007-6157-7398]{Lauren A. Elicker}
\affiliation{Department of Physics and Astronomy and PITT PACC, University of Pittsburgh, Pittsburgh, PA 15260, USA}

\author[0000-0002-2862-307X]{M. Riley Owens}
\affiliation{Department of Astronomy, University of California, Berkeley, Berkeley, CA 94720, USA}

\author[0009-0003-3123-4897]{Prasanna Adhikari}
\affiliation{Department of Physics, University of Cincinnati, Cincinnati, OH 45221, USA}

\author[0000-0002-3475-7648]{Gourav Khullar}
\affiliation{Department of Astronomy, University of Washington, Seattle, WA 98195, USA}

\begin{abstract}

Strong lensing galaxy clusters provide a unique and powerful way to test simulation-derived structure predictions that follow from $\Lambda$ Cold Dark Matter ($\Lambda$CDM) cosmology. Specifically, the relative alignments of the dark matter (DM) halo, stars, and hot intracluster gas in these clusters offer insights into how well theoretical structure predictions hold. We measure the position angles, ellipticities, and locations/centroids of the brightest cluster galaxy (BCG), the Intracluster Light (ICL), the hot Intracluster Medium (ICM), and the Core Lensing Mass (CLM) for a sample of strong lensing galaxy clusters from the Sloan Giant Arcs Survey (SGAS). We measure the shapes (position angles and ellipticities) and centroids of these distributions using ellipse-fitting methods applied to different datasets: \emph{HST} WFC3 imaging for the BCG and ICL, \emph{Chandra} X-ray observations for the ICM, and strong-lensing mass reconstructions for the CLM. Additionally, we incorporate ICM morphological measures to classify the dynamical state of the cluster sample. Using this multi-component approach, we constrain the shape and centroids of these distributions in this sample and evaluate the different observable components in terms of their ability to trace the gravitational potential of their respective clusters. We find that misalignments between cluster components can be explained by astrophysical processes related to cluster assembly, relaxation, and merger histories. We find that the ICL is most closely aligned with its host DM halo, as traced by the CLM distribution, in both position angle and centroid. Additionally, we find that on average the ICL and CLM are more elliptical than the ICM and BCG.

\end{abstract}


\section{Introduction}
\label{sec:intro}
\subsection{Galaxy Clusters in $\Lambda$CDM}

\begin{figure*}
	\includegraphics[width=\textwidth]{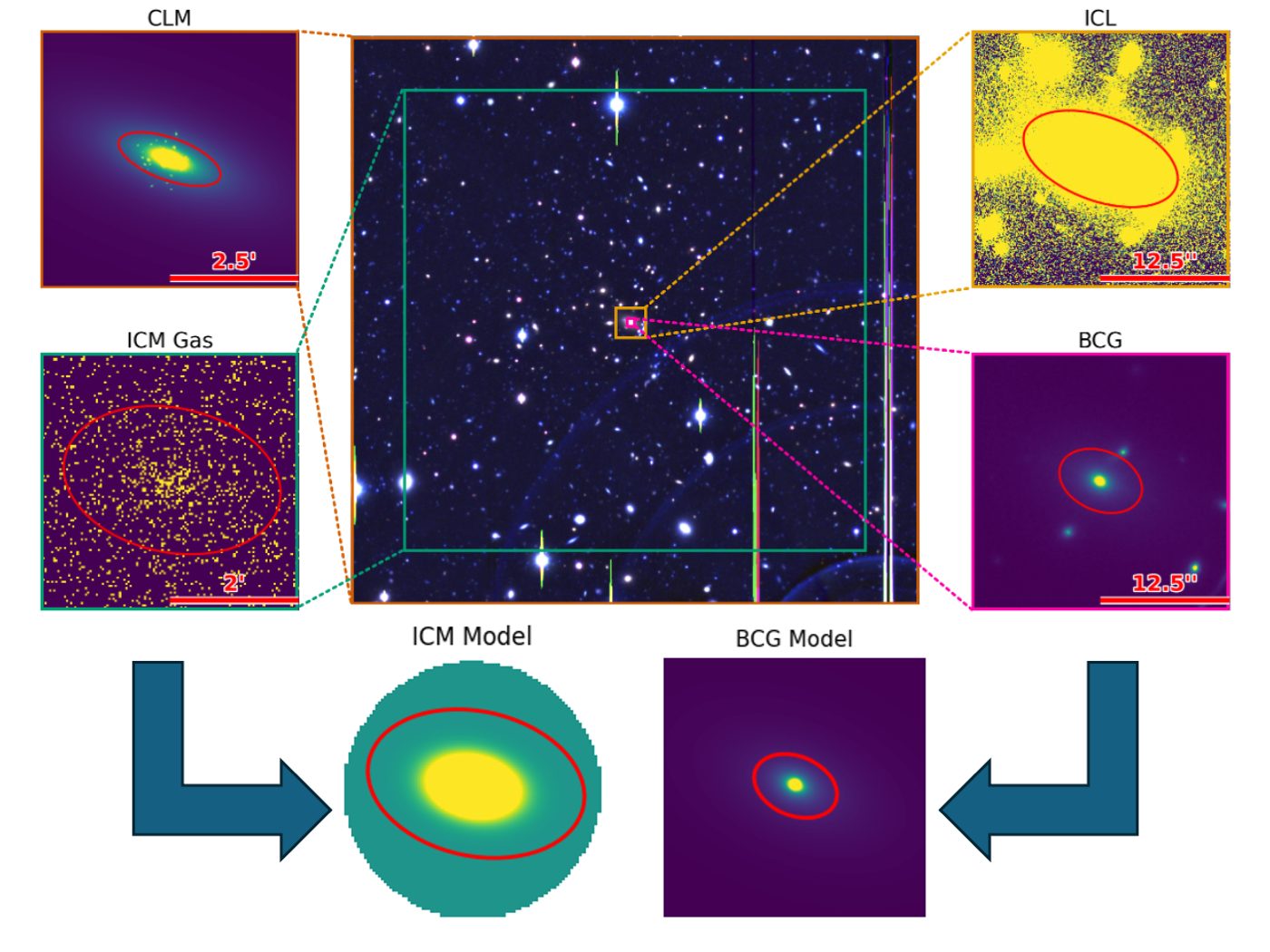}
    \caption{Visualization of the cluster components considered in this work (the CLM, ICL, BCG, and ICM) for an example galaxy cluster, J0957+0509. Each panel displays a red ellipse representing the measured shape and centroid. Center panel: a $5''\times5''$ representative color (gri) Subaru image of the field \citep{Oguri2012} for a scale comparison of the different cluster components. Top-left: the minimum $\chi^2$ mass map generated from the lensing profile derived in \citet{Sharon2020}. Top-right: \emph{HST} WFC3/IR F160W imaging data \citep{Sharon2020} tuned to highlight the ICL distribution. Bottom-Left: \emph{Chandra} data displaying the ICM \citep{Gassis2025} with an arrow pointing to the ICM model derived using \texttt{CIAO Sherpa}. Bottom-Right: \emph{HST} WFC3/IR F160W imaging data \citep{Sharon2020} tuned to highlight the BCG distribution with an arrow pointing to the BCG model derived using \texttt{GALFIT} \citep{Peng2002}.}
    \label{fig:data}
\end{figure*}

An important consequence of $\Lambda$ Cold Dark Matter ($\Lambda$CDM) physics is the concept that galaxy clusters are built up by the hierarchical formation of formerly separate dark matter (DM) halo systems. In this scenario, the cluster gradually accretes and incorporates other halos, eventually forming one large halo system that defines the cluster \citep*{Beers1983}. Due to this process of hierarchical mergers, clusters are the most massive self-gravitating objects in the known universe \citep{Djorgovski1987}.

DM comprises $\sim$85$\%$ of the mass in a given galaxy cluster \citep{plank2016}. The largest baryonic component is a hot, diffuse, X-ray–emitting plasma that permeates the cluster, known as the Intracluster Medium (ICM) \citep{Sarazin1986}. The remaining baryons lie in several stellar components: (1) the uniquely massive and centrally located galaxy known as brightest cluster galaxy (BCG) \citep{Sastry1968}; (2) the diffuse stellar component stripped from galaxies that subtends the cluster and is built up around the BCG known as the Intracluster Light (ICL) \citep{Montes2018}; and (3) the cluster member galaxies that orbit within the cluster potential \citep{Dressler1980}.

In a theoretical system only subject to gravitational forces, the BCG, ICL, and ICM of a given galaxy cluster should align with the DM halo of the cluster \citep{Sastry1968,Binggeli1982,Hashimoto2008,Niederste-Ostholt2010,Biernacka2015,Donahue2016,Wang2018}. Even when we incorporate all aspects of the complex astrophysical landscape of these galaxy clusters, we still expect the differences between the position angles, ellipticities, and centroids of these distributions to be small and infrequent.

\subsection{Structure Alignment Hypotheses for the Mass Components of Galaxy Clusters}

As a direct consequence of the hierarchical formation of galaxy clusters via halo mergers, we would expect that once the galaxy cluster has formed and relaxed, the BCG should reside at the gravitational center of the cluster potential \citep{West1994,Dubinski1998,Donahue2016,Okabe2020b,Ragone-Figueroa2020}. This is because the BCG assembles primarily from the stellar mass of the most massive progenitor galaxies in the merging halos, and we expect the stellar and DM components of each halo to align. In addition, we expect the BCG to share its orientation with its DM halo due to mergers happening preferentially along the major axis of the DM halo \citep{Sastry1968,Binggeli1982,Niederste-Ostholt2010,Biernacka2015}.

We would also expect the non-galactic stellar matter component, the ICL, to share its centroid with the DM halo as well. This is because the most dominant channels by which the ICL forms are mergers of central galaxies with satellite galaxies (\citealt*{Gregg1998}; \citealt{Mihos2005,Contini2014}; \citealt*{Montes2014}; \citealt{Groenewald2017,Contini2018}; \citealt*{Montes2018}), tidal stripping of stars from intermediate and massive galaxies (\citealt{Rudick2009}; \citealt*{Martel2012}; \citealt{Contini2014,Contini2018}), and pre-processing/accretion (\citealt*{Rudick2006}; \citealt{Larsen2006,Contini2014,Contini2018}). Though these processes can happen anywhere in the cluster, they are concentrated around the gravitational center of the halo, coincident with the BCG since there is a higher mass concentration, satellite galaxy density, and merger frequency in the core regions of the cluster. Additionally, the ICL created in all areas should eventually spread out to trace the DM halo distribution and gravitational gradient, since the ICL is gravitationally bound to the cluster as opposed to any individual cluster member \citep{Pillepich2014,Montes2018,Pillepich2018}.

The ICM should also trace the gravitational potential of the cluster because it is also dominantly subject to the gravitational forces of the DM halo as opposed to the individual cluster members (\citealt*{Forman1982}; \citealt{Donahue2016}). For relaxed clusters, the gravitational potential and entropy of the gas determine the ICM's spatial distribution \citep{Voit2001}. Thus, X-ray observations of the ICM should trace the shape, centroid, and slope of the DM dominated gravitational potential of the cluster \citep{Wik2008}.

The DM halo itself must be well constrained such that we can compare the observed baryonic components to the cluster's main halo. By using a sample of strong lensing galaxy clusters, we can study the small-scale structure of the DM distribution of the lensing cluster to a great degree of precision. Effectively, we are able to make a direct measurement of the projected DM distribution of a galaxy cluster using well-constrained lensing profiles \citep[e.g.,][]{Merten2015,Zitrin2015,Jauzac2016}. These lensing profiles allow us to reconstruct the projected, DM dominated mass distribution of the cluster which we refer to as the Core Lensing Mass (from here on, CLM). The CLM provides a robust tracer of the inner DM halo structure and serves as a reference for comparison to other cluster observables.

\subsection{Misalignments and their Physical Significance}

Previous work has found that sometimes the various mass components are not aligned with the DM halo or with each other in regards to position angle, ellipticity, and/or centroid (e.g., \citealt{Bosch2005}; \citealt*{Sanderson2009}; \citealt{Skibba2011,Zitrin2012,Hikage2013,Lauer2014}; \citealt*{Martel2014}; \citealt{Oliva-Altamirano2014,Wang2014,Hoshino2015,Rossetti2016,Lange2018,Lopes2018,Zenteno2020,Propis2021}). Consequently, this suggests that there are mechanisms within or beyond the context of $\Lambda$CDM by which these misalignments can occur. 

If we observe a misalignment between the BCG and CLM, it could suggest that a major merger recently occurred. This is because we predict that given enough time, the BCG should revert to its shared orientation with the DM halo \citep{Wittman2019}. In this way, the BCG should serve as a tracer of the merger history of the galaxy cluster. However, work by \citet{Kim2017} and \citet{Harvey2017} have found that the BCG can still show misalignments in relaxed clusters. This implies that the BCG “wobbles” after cluster relaxation has been achieved. This could suggest that a beyond-standard DM model may better describe the underlying physics than the conventional CDM framework. Alternatively, more recent work by \citet{Roche2024} showed that these wobbles may be a consequence of lower simulation resolution. In which case, the classic CDM explanation may remain the most viable option when accounting for systematic inflation of BCG and CLM offsets.

\citet{Montes2018} claim that the ICL should closely trace the DM halo distribution. Even in disturbed clusters, the ICL should still align with the CLM. Expanding ICL tests to a larger cluster sample in various dynamical states will allow us to quantify the degree to which the ICL functions as a tracer for DM and investigate potential mechanisms that may drive misalignments between the two distributions. 

The ICM can exhibit extreme misalignments for disturbed clusters due to merger activity. However, the ICM can also show misalignments even for relaxed clusters. This is a consequence of hydrodynamical gas oscillations that occur as a result of previous major mergers. This sloshing can persist even when the other components of the cluster system have had time to relax with respect to one another (\citealt*{Markevitch2001}; \citealt{Churazov2003,Johnson2012,Harvey2017}).

\subsection{Paper Outline}

This paper constrains the degree to which the BCG, ICL, and ICM deviate from alignment with respect to their DM halo counterpart characterized by the CLM for a sample of strong lensing clusters over a range of redshifts ($0.176<z<0.656$) at different phases of dynamical relaxation. We do this by fitting ellipses to the aforementioned components. This allows us to measure the position angles, ellipticities, and centroids of the various components of the galaxy clusters in this sample and compare them to each other. From this, we will be able to constrain how often deviations occur and what mechanisms result in misalignments between the various components. In Section~\ref{sec:dat-meas}, we describe the cluster sample, highlighting the data and methods used to make measurements on each of the components introduced in this Section~\ref{sec:intro}. In Section~\ref{sec:results}, we present results, specifically comparing the differences between the observables we measure from our ellipse-fitting methods. In Section~\ref{sec:disc}, we discuss in more detail the correlations between different measurements, as well as the impact of dynamical state, allowing us to constrain the physical processes that may cause misalignment. In this paper, we assume a flat $\Lambda$CDM cosmology with $\Omega_\Lambda = 0.7$, $\Omega_m = 0.3$, and $H_0 = 70$ $\mathrm{km}$ $\mathrm{s}^{-1}$ $\mathrm{Mpc}^{-1}$.

\section{Data and Measurements}
\label{sec:dat-meas}

\subsection{Cluster Sample}
\label{sec:clus-samp} 

We use a sample of 39 strong lensing galaxy clusters to measure the shapes and centroids of the BCG, ICL, and strong lensing derived CLM. All clusters with the exception of PSZ1G311 \citep{Sharon2022b} are part of the Sloan Giant Arcs Survey (SGAS) \citep{koester2010,bayliss2011,bayliss2014,gladders2013,Sharon2020} cluster sample. Each cluster has multi-band \emph{HST} WFC3 imaging data and spectroscopic follow-up that enabled strong gravitational lensing models \citep{Sharon2020,Sharon2022a,Sharon2022b}. The sample consists of clusters selected solely for their bright, strongly lensed background galaxies. A subset of 27 clusters have \emph{Chandra} ACIS-I/S X-ray data \citep{Gassis2025} which allows us to measure the ICM in tandem with the other three components. Figure~\ref{fig:data} illustrates the BCG, ICL, ICM, and CLM for an example galaxy cluster with ellipses overlaid representing the measured shape and centroid parameters described in Sections~\ref{sec:hst}-\ref{sec:clm}.

\subsubsection{Double-Cored Systems} \label{sec:dc_sys}

For most of the clusters, the ICM and CLM correspond to a single optical core system composed of the BCG and the surrounding ICL. We find three systems in which two optical cores are very close together with respect to redshift and angular separation. We discuss these three systems in detail since they deviate from the regular one-to-one matching of the cluster components considered in this paper.

For J0928+2031, a singular distribution characterized by the North-Western core overtly dominates the lensing mass map. The secondary, South-Eastern core is very subdominant, comparable to other features in the mass map. Though identifiable, the secondary core is also subdominant compared to the primary core in the X-ray data. Although there appear to be two optical cores, one dominant feature best describes the CLM and large-scale ICM. For the purpose of this study, it is most consistent to measure the two BCG+ICL systems separately but compare them to the same singular, very dominant ICM and CLM distributions.

For J2243$-$0935, the two cores are comparable in size in the strong lensing mass maps. Though one is larger than the other, they both clearly stand out from other features on the projected mass map. In the \emph{Chandra} data, the ICM distribution appears as a merging system represented by a singular elliptical shape rather than two distinguishable structures. As such, we measure the ICL, BCG, and CLM for the Eastern and Western core separately but compare them to the same singular ICM distribution. 

For J1226+2149 and J1226+2152, all components are completely distinguishable, though they exist very close together in terms of redshift and position on the sky. Though beyond the scope of this paper, we note that this system may be in a pre-merger state or in the very early stages of its merging process (with an additional third BCG+ICL system outside the sample of this paper). In consequence, we measure ICL, BCG, ICM, and CLM completely separately, treating J1226+2149 and J1226+2152 as two fully separate systems for the purpose of this paper.

\begin{deluxetable*}{ccccccccc}
\centerwidetable
\setlength{\tabcolsep}{6pt}
\def\arraystretch{1.15}
\tablecaption{This table lists the measured elliptical properties of the Brightest Cluster Galaxy (BCG) and Intracluster Light (ICL). We report the right ascension and declination in degrees (J2000). We report the position angle in degrees, measured counterclockwise from West ($0^\circ$ corresponds to West). We define the reported ellipticity according to equation~\ref{eq:eps}.\label{tab:bcg+icl_meas}}
\tablehead{
\colhead{Name} & \colhead{$\alpha_{BCG}$} & \colhead{$\delta_{BCG}$} & \colhead{PA$_{BCG}$} & \colhead{$\emph{e}_{BCG}$} & \colhead{$\alpha_{ICL}$} & \colhead{$\delta_{ICL}$} & \colhead{PA$_{ICL}$} & \colhead{$\emph{e}_{ICL}$}
}
\startdata
J0004$-$0103 & 1.21642 & -1.05440 & $41^{+36}_{-1}$ & $0.26^{+0.01}_{-0.19}$ & 1.21576 & -1.05501 & $133^{+1}_{-0.9}$ & $0.53^{+0.03}_{-0.04}$ \\
J0108+0623 & 17.17513 & 6.41210 & $158^{+3}_{-2}$ & $0.06^{+0.09}_{-0.01}$ & 17.17506 & 6.41204 & $156^{+0.9}_{-4}$ & $0.37^{+0.03}_{-0.06}$ \\
J0146$-$0929 & 26.73335 & -9.49792 & $69^{+87}_{-3}$ & $0.05^{+0.02}_{-0.03}$ & 26.73351 & -9.49773 & $128^{+10}_{-9}$ & $0.23^{+0.03}_{-0.04}$ \\
J0150+2725 & 27.50357 & 27.42675 & $48.4^{+0.1}_{-0.3}$ & $0.23^{+0.02}_{-0.04}$ & 27.50351 & 27.42675 & $47^{+5}_{-6}$ & $0.3^{+0.06}_{-0.03}$ \\
J0333$-$0651 & 53.26955 & -6.85625 & $1^{+1}_{-3}$ & $0.08^{+0.06}_{-0.02}$ & 53.2695 & -6.85643 & $175^{+3}_{-9}$ & $0.11^{+0.04}_{-0.03}$ \\
J0851+3331 & 132.91194 & 33.51837 & $44.0^{+0.4}_{-0.6}$ & $0.27^{+0.05}_{-0.02}$ & 132.91231 & 33.51935 & $49^{+27}_{-2}$ & $0.43^{+0.06}_{-0.07}$ \\
J0915+3826 & 138.91255 & 38.44961 & $59^{+16}_{-2}$ & $0.36^{+0.13}_{-0.13}$ & 138.91391 & 38.44935 & $13^{+1}_{-2}$ & $0.35^{+0.13}_{-0.007}$ \\
J0928+2031 (NW) & 142.01891 & 20.52919 & $51^{+4}_{-20}$ & $0.08^{+0.05}_{-0.04}$ & 142.01911 & 20.52898 & $62^{+2}_{-2}$ & $0.27^{+0.05}_{-0.02}$ \\
J0928+2031 (SE) & 142.02738 & 20.51900 & $121^{+4}_{-2}$ & $0.16^{+0.04}_{-0.05}$ & 142.02983$^*$ & 20.51889$^*$ & $108^{+1}_{-0.9}$ & $0.38^{+0.02}_{-0.01}$ \\
J0952+3434 & 148.16761 & 34.57948 & $81^{+3}_{-4}$ & $0.22^{+0.09}_{-0.06}$ & 148.16728 & 34.57946 & $85^{+34}_{-4}$ & $0.29^{+0.05}_{-0.12}$ \\
J0957+0509 & 149.41329 & 5.15884 & $157.4^{+0.7}_{-0.5}$ & $0.3^{+0.09}_{-0.02}$ & 149.41314 & 5.15875 & $160^{+2}_{-0.9}$ & $0.48^{+0.02}_{-0.03}$ \\
J1038+4849 & 159.68158 & 48.82160 & $21.7^{+0.5}_{-0.4}$ & $0.15^{+0.05}_{-0.04}$ & 159.67994 & 48.82181 & $21^{+0.9}_{-1}$ & $0.5^{+0.07}_{-0.02}$ \\
J1050+0017 & 162.66625 & 0.28533 & $93^{+8}_{-21}$ & $0.04^{+0.01}_{-0.009}$ & 162.66627 & 0.28546 & $86^{+8}_{-11}$ & $0.12^{+0.05}_{-0.04}$ \\
J1055+5547 & 163.76919 & 55.80648 & $89.9^{+0.2}_{-0.5}$ & $0.18^{+0.02}_{-0.01}$ & 163.77002 & 55.80622 & $78^{+31}_{-4}$ & $0.40^{+0.02}_{-0.27}$ \\
J1110+6459 & 167.57379 & 64.99665 & $151^{+5}_{-7}$ & $0.3^{+0.02}_{-0.005}$ & 167.57385$^*$ & 64.99667$^*$ & $158^{+2}_{-3}$ & $0.48^{+0.04}_{-0.14}$ \\
J1115+1645 & 168.76831 & 16.76072 & $83^{+5}_{-0.9}$ & $0.33^{+0.10}_{-0.03}$ & 168.76841 & 16.76056 & $92^{+3}_{-0.9}$ & $0.55^{+0.02}_{-0.02}$ \\
J1138+2754 & 174.53732 & 27.90854 & $82^{+2}_{-1}$ & $0.28^{+0.05}_{-0.006}$ & 174.53738 & 27.90858 & $84^{+1}_{-4}$ & $0.64^{+0.02}_{-0.1}$ \\
J1152+3312 & 178.00078 & 33.22826 & $176^{+0.6}_{-2}$ & $0.21^{+0.08}_{-0.07}$ & 178.00137 & 33.22834 & $174^{+2}_{-2}$ & $0.35^{+0.05}_{-0.06}$ \\
J1152+0930 & 178.19748 & 9.50410 & $57^{+2}_{-0.5}$ & $0.09^{+0.07}_{-0.03}$ & 178.1972 & 9.50415 & $54^{+4}_{-16}$ & $0.42^{+0.07}_{-0.15}$ \\
J1156+1911 & 179.02279 & 19.18685 & $152^{+15}_{-49}$ & $0.11^{+0.03}_{-0.04}$ & 179.02268 & 19.18662 & $106^{+3}_{-2}$ & $0.36^{+0.02}_{-0.04}$ \\
J1207+5254 & 181.89964 & 52.91645 & $31^{+2}_{-2}$ & $0.13^{+0.01}_{-0.02}$ & 181.90096 & 52.91752 & $119.6^{+0.7}_{-0.8}$ & $0.51^{+0.02}_{-0.02}$ \\
J1209+2640 & 182.34864 & 26.67961 & $143^{+1}_{-0.9}$ & $0.17^{+0.03}_{-0.02}$ & 182.34813 & 26.67904 & $157^{+2}_{-4}$ & $0.32^{+0.06}_{-0.06}$ \\
J1226+2149 & 186.71298 & 21.83120 & $46^{+1}_{-0.8}$ & $0.132^{+0.005}_{-0.003}$ & 186.71284 & 21.83124 & $61^{+1}_{-2}$ & $0.14^{+0.009}_{-0.02}$ \\
J1226+2152 & 186.71541 & 21.87372 & $114.6^{+0.8}_{-0.5}$ & $0.27^{+0.05}_{-0.02}$ & 186.71535 & 21.87373 & $114^{+5}_{-0.8}$ & $0.48^{+0.04}_{-0.009}$ \\
J1329+2243 & 202.39392 & 22.72106 & $121^{+2}_{-0.5}$ & $0.055^{+0.004}_{-0.003}$ & 202.39413 & 22.72099 & $180^{+0.9}_{-3}$ & $0.19^{+0.02}_{-0.008}$ \\
J1336$-$0331 & 204.00041 & -3.52500 & $46^{+0.9}_{-2}$ & $0.18^{+0.03}_{-0.02}$ & 204.00071$^*$ & -3.52511$^*$ & $45^{+72}_{-7}$ & $0.22^{+0.06}_{-0.13}$ \\
J1343+4155 & 205.88685 & 41.91762 & $153.1^{+0.5}_{-0.9}$ & $0.26^{+0.09}_{-0.07}$ & 205.88757$^*$ & 41.91766$^*$ & $157^{+3}_{-1}$ & $0.46^{+0.02}_{-0.02}$ \\
J1420+3955 & 215.16798 & 39.91924 & $56^{+14}_{-7}$ & $0.07^{+0.04}_{-0.03}$ & 215.16806 & 39.9195 & $118^{+2}_{-12}$ & $0.34^{+0.15}_{-0.06}$ \\
J1429+1202 & 217.47838 & 12.04314 & $66^{+0.9}_{-2}$ & $0.09^{+0.01}_{-0.009}$ & 217.47835 & 12.04326 & $66^{+1}_{-1}$ & $0.138^{+0.006}_{-0.004}$ \\
J1439+1208 & 219.79075 & 12.14043 & $7^{+8}_{-1}$ & $0.23^{+0.03}_{-0.04}$ & 219.79053 & 12.14055 & $38^{+3}_{-8}$ & $0.28^{+0.03}_{-0.01}$ \\
J1456+5702 & 224.00359 & 57.03903 & $24^{+6}_{-16}$ & $0.026^{+0.004}_{-0.008}$ & 224.00344 & 57.03913 & $59^{+11}_{-5}$ & $0.19^{+0.08}_{-0.01}$ \\
J1522+2535 & 230.71986 & 25.59097 & $12.1^{+0.9}_{-0.2}$ & $0.15^{+0.07}_{-0.03}$ & 230.72004 & 25.59113 & $14^{+10}_{-2}$ & $0.34^{+0.02}_{-0.02}$ \\
J1531+3414 & 232.79423 & 34.24019 & $41^{+43}_{-4}$ & $0.2^{+0.02}_{-0.08}$ & 232.79387 & 34.24016 & $119^{+20}_{-9}$ & $0.19^{+0.02}_{-0.06}$ \\
PSZ1G311 & 237.52959 & -78.19169 & $104^{+6}_{-8}$ & $0.049^{+0.009}_{-0.004}$ & 237.52974 & -78.19162 & $103^{+5}_{-9}$ & $0.11^{+0.06}_{-0.03}$ \\
J1604+2244 & 241.04228 & 22.73858 & $139^{+3}_{-5}$ & $0.156^{+0.007}_{-0.009}$ & 241.04241 & 22.7379 & $145^{+3}_{-63}$ & $0.2^{+0.14}_{-0.08}$ \\
J1621+0607 & 245.38488 & 6.12200 & $6^{+39}_{-5}$ & $0.03^{+0.01}_{-0.009}$ & 245.38497 & 6.12189 & $53^{+20}_{-11}$ & $0.29^{+0.03}_{-0.08}$ \\
J1632+3500 & 248.04523 & 35.00967 & $169^{+2}_{-3}$ & $0.19^{+0.04}_{-0.01}$ & 248.04433 & 35.0091 & $155^{+0.8}_{-1}$ & $0.5^{+0.02}_{-0.05}$ \\
J1723+3411 & 260.90064 & 34.19948 & $173^{+2}_{-0.6}$ & $0.13^{+0.04}_{-0.007}$ & 260.90063 & 34.19946 & $176^{+3}_{-2}$ & $0.25^{+0.08}_{-0.03}$ \\
J2111$-$0114 & 317.83063 & -1.23985 & $122^{+3}_{-21}$ & $0.051^{+0.009}_{-0.002}$ & 317.83079 & -1.23978 & $76^{+7}_{-11}$ & $0.12^{+0.01}_{-0.009}$ \\
J2243$-$0935 (W) & 340.83636 & -9.58859 & $1.0^{+0.5}_{-0.1}$ & $0.18^{+0.02}_{-0.01}$ & 340.83582 & -9.58911 & $137^{+6}_{-0.8}$ & $0.63^{+0.03}_{-0.16}$ \\
J2243$-$0935 (E) & 340.85546 & -9.58422 & $68.8^{+0.4}_{-0.4}$ & $0.080^{+0.002}_{-0.002}$ & 340.85503 & -9.58269 & $103^{+1}_{-3}$ & $0.33^{+0.09}_{-0.02}$ \\
\enddata
\tablecomments{We omit errors on $\alpha$ and $\delta$ since they are determined to within $< 1''$ precision for the majority of BCG and ICL measurements. The $*$ marker indicates a measured $\alpha$/$\delta$ with a slightly higher uncertainty ($<1.8''$).}
\end{deluxetable*}

\subsection{\emph{HST} Data}
\label{sec:hst}
For measurements of the BCG and ICL components of the SGAS cluster sample, we use \emph{HST} WFC3 imaging data. The majority of observations were part of the SGAS-HST program (GO-13003, PI: Gladders) \citep[for details, see][]{Sharon2020}. We prioritize \emph{HST} WFC3/IR F160W imaging data when making measurements of the BCG and ICL, since this is the reddest filter available for the clusters in this program. For some clusters, we do not have \emph{HST} WFC3/IR F160W imaging data from the SGAS-HST program. J1429+1202 and J1226+2152 have \emph{HST} WFC3/IR F160W imaging from GO-15378 (PI: Bayliss), and PSZ1G311 has \emph{HST} WFC3/IR F160W imaging from GO-15377 (PI: Bayliss). J1226+2149 is a unique case where our preferred filter was unavailable; instead, we use the reddest filter available from GO-12368 (PI: Morris): \emph{HST} WFC3/UVIS F606W. We see in Appendix \ref{appendix:bcg+icl} that the ICL and BCG components for J1226+2149 are detectable and comparable in quality to our \emph{HST} WFC3/IR F160W measurements. The measured right ascension ($\alpha$), declination ($\delta$), position angle (PA), and ellipticity (\emph{e}) are listed in Table~\ref{tab:bcg+icl_meas} for each BCG and ICL distribution.

\subsubsection{BCG Measurement}
\label{sec:bcg}

Before making measurements of the BCG, we first identified it in each cluster. In most cases, the BCG was unambiguous; it appeared as a distinctly bright elliptical galaxy, situated near the centroid of the strong-lensing arcs and the surrounding ICL.

In the few cases where the BCG could not be unambiguously identified from single-band imaging data alone, we selected the BCG using a color–magnitude and color–color analysis of the cluster field using publicly available SDSS photometry. Candidates were objects that (1) lie within 0.15 magnitudes of the cluster red sequence in the SDSS r–band and i–band color–magnitude space, (2) fall within 0.15 magnitudes of the over-density of red-sequence galaxies identified in the g–r versus r–i color–color diagram, using the SDSS g, r, and i filters, and (3) are spatially located within the region enclosed by the radius of curvature of the strong-lensing arcs. Among these candidates, the brightest galaxy was selected as the BCG, which was confirmed by visual inspection. In a few cases, this change in method gives us different measurements than those published in the \citet{Gassis2024} conference proceedings, which selected the galaxy closest to the gravitational center of the cluster determined by the CLM distribution center to be the BCG. The difference in BCG selection does not affect the qualitative conclusions between these two papers.

In this sample, there are active galaxy-galaxy mergers occurring which complicate BCG selection. In cases where the BCG consists of multiple merging galaxies, we define the BCG as the collection of all objects located within 5 kpc of the peak brightness of the identified system.

After we identified the BCG, we needed to isolate it from all other objects in the field to exclude any interfering light from the measurement. To do this, we modeled the BCGs using \texttt{GALFIT} \citep{Peng2002}. In most cases, double Sérsic models \citep{Sersic1968} were required to accurately model the BCG. We allowed the parameters of both components to freely explore the parameter space when iterating through \texttt{GALFIT} with minor constraints on centroid and radius to prevent the outer Sérsic from fitting to the ICL. 

Using the isolated BCG models outputted from \texttt{GALFIT}, we measured the elliptical shape using the \texttt{lsq-ellipse}\footnote{\url{https://github.com/bdhammel/least-squares-ellipse-fitting}; \citealt{Hammel2020}} Python package which employs the method of least squares \citep{Halir1998} to get accurate ellipse estimates. Since we used double Sérsic models that span a large spatial regime, it was difficult to differentiate where the BCG ended and the ICL began. We measure the BCG elliptical properties from our selected transition radius of 30 kpc down to an inner limit of 5 kpc, below which we observe artificial sphericalization of the BCG in some systems due to pixel saturation and complications due to galaxy-galaxy mergers in the core. 

\subsubsection{ICL Measurement}
\label{sec:icl}

For the measurements of the ICL distribution, we use iterative elliptical isophote fitting with masks. From study-to-study, the definition and observability of the ICL can vary. For our purposes, we take the ICL to be the light that envelops the core of the cluster that extends beyond the gravitational influence of the BCG. 

Therefore, we need to mask out all the light contributions from non-BCG galaxy members for which the ICL envelops and measure beyond the BCG-ICL transition radius \citep{Contini2022}. To isolate the ICL, we separated it from all other objects in the field using SExtractor to derive masks \citep{Bertin1996}. We allowed objects to be masked at 2.5 or 3.5 times the Kron radius \citep{Kron1980} depending on what fit the object more optimally. We additionally masked lensed objects and stars that are not precisely identified by the SExtractor code. 

Once the ICL was isolated, we used iterative isophote fitting applying the python \texttt{lsq-ellipse} function to measure the ICL out to the largest radii. Effectively, we find the isophote value for which the ICL still remains distinguishable from the background and iteratively measure the ICL until we reach our selected BCG-ICL transition radius of 30 kpc \citep{Kluge2021}.

\begin{deluxetable*}{ccccc}
\centerwidetable
\setlength{\tabcolsep}{12pt}
\def\arraystretch{1.2}
\tablecaption{This table lists the measured elliptical properties of the Intracluster Medium (ICM) We report the right ascension and declination in degrees (J2000). We report the position angle in degrees, measured counterclockwise from West ($0^\circ$ corresponds to West). We define the reported ellipticity according to equation~\ref{eq:eps}.\label{tab:icm_meas}}
\tablehead{
\colhead{Name} & \colhead{$\alpha_{ICM}$} & \colhead{$\delta_{ICM}$} & \colhead{PA$_{ICM}$} & \colhead{$\emph{e}_{ICM}$}
}
\startdata
J0851+3331 & $132.909^{+0.002}_{-0.0006}$ & $33.521^{+0.001}_{-0.0005}$ & $24^{+7}_{-6}$ & $0.34^{+0.1}_{-0.06}$ \\
J0928+2031 (NW) & $142.0181^{+0.0003}_{-0.0003}$ & $20.5269^{+0.0003}_{-0.0003}$ & $62^{+10}_{-11}$ & $0.11^{+0.04}_{-0.04}$ \\
J0928+2031 (SE) & $142.0181^{+0.0003}_{-0.0003}$ & $20.5269^{+0.0003}_{-0.0003}$ & $62^{+10}_{-11}$ & $0.11^{+0.04}_{-0.04}$ \\
J0952+3434 & $148.1679^{+0.0005}_{-0.0005}$ & $34.5771^{+0.0006}_{-0.0006}$ & $98^{+6}_{-6}$ & $0.40^{+0.05}_{-0.06}$ \\
J0957+0509 & $149.4142^{+0.0006}_{-0.0006}$ & $5.1592^{+0.0004}_{-0.0004}$ & $168^{+12}_{-11}$ & $0.34^{+0.1}_{-0.12}$ \\
J1038+4849 & $159.6792^{+0.0003}_{-0.0003}$ & $48.822^{+0.0002}_{-0.0002}$ & $163^{+26}_{-28}$ & $0.07^{+0.05}_{-0.06}$ \\
J1050+0017 & $162.6667^{+0.0003}_{-0.0003}$ & $0.2858^{+0.0004}_{-0.0004}$ & $54^{+10}_{-11}$ & $0.19^{+0.06}_{-0.07}$ \\
J1110+6459 & $167.5742^{+0.0009}_{-0.0009}$ & $64.9975^{+0.0004}_{-0.0004}$ & $130^{+7}_{-7}$ & $0.30^{+0.07}_{-0.07}$ \\
J1138+2754 & $174.5381^{+0.0003}_{-0.0003}$ & $27.9059^{+0.0004}_{-0.0004}$ & $92^{+4}_{-4}$ & $0.33^{+0.04}_{-0.04}$ \\
J1152+3312 & $178.003^{+0.001}_{-0.002}$ & $33.2324^{+0.0007}_{-0.0009}$ & $153^{+28}_{-19}$ & $0.22^{+0.13}_{-0.14}$ \\
J1152+0930 & $178.1982^{+0.0005}_{-0.0005}$ & $9.5039^{+0.0006}_{-0.0006}$ & $63^{+12}_{-13}$ & $0.24^{+0.06}_{-0.09}$ \\
J1207+5254 & $181.8948^{+0.0008}_{-0.0007}$ & $52.9127^{+0.0007}_{-0.0005}$ & $118^{+5}_{-4}$ & $0.36^{+0.06}_{-0.04}$ \\
J1209+2640 & $182.3476^{+0.0004}_{-0.0003}$ & $26.6782^{+0.0002}_{-0.0002}$ & $158^{+9}_{-10}$ & $0.17^{+0.05}_{-0.06}$ \\
J1226+2149 & $186.7131^{+0.0002}_{-0.0002}$ & $21.8314^{+0.0002}_{-0.0002}$ & $18^{+5}_{-4}$ & $0.16^{+0.02}_{-0.02}$ \\
J1226+2152 & $186.7156^{+0.0002}_{-0.0002}$ & $21.8732^{+0.0002}_{-0.0002}$ & $110^{+4}_{-4}$ & $0.29^{+0.04}_{-0.04}$ \\
J1329+2243 & $202.3938^{+0.0005}_{-0.0004}$ & $22.7229^{+0.0005}_{-0.0005}$ & $101^{+14}_{-14}$ & $0.17^{+0.07}_{-0.07}$ \\
J1336$-$0331 & $204.0013^{+0.0004}_{-0.0004}$ & $-3.5245^{+0.0004}_{-0.0004}$ & $57^{+10}_{-11}$ & $0.14^{+0.04}_{-0.05}$ \\
J1343+4155 & $205.8869^{+0.0003}_{-0.0003}$ & $41.9176^{+0.0002}_{-0.0002}$ & $150^{+4}_{-4}$ & $0.3^{+0.04}_{-0.04}$ \\
J1420+3955 & $215.1685^{+0.0004}_{-0.0004}$ & $39.9198^{+0.0003}_{-0.0003}$ & $0^{+1}_{-1}$ & $0.04^{+0.07}_{-0.01}$ \\
J1439+1208 & $219.7906^{+0.0003}_{-0.0003}$ & $12.1415^{+0.0002}_{-0.0002}$ & $24^{+12}_{-12}$ & $0.14^{+0.05}_{-0.05}$ \\
J1456+5702 & $224.0025^{+0.001}_{-0.0008}$ & $57.0393^{+0.0005}_{-0.0005}$ & $25^{+1}_{-1}$ & $0.05^{+0.11}_{-0.01}$ \\
J1522+2535 & $230.7195^{+0.0005}_{-0.0005}$ & $25.5932^{+0.0004}_{-0.0004}$ & $156^{+1}_{-1}$ & $0.10^{+0.12}_{-0.01}$ \\
J1531+3414 & $232.7942^{+9e-05}_{-0.0001}$ & $34.24008^{+8e-05}_{-9e-05}$ & $46^{+17}_{-18}$ & $0.09^{+0.06}_{-0.05}$ \\
PSZ1G311 & $237.5222^{+0.0008}_{-0.0008}$ & $-78.1911^{+0.0002}_{-0.0002}$ & $40^{+8}_{-8}$ & $0.12^{+0.03}_{-0.03}$ \\
J1621+0607 & $245.3835^{+0.0008}_{-0.0008}$ & $6.123^{+0.001}_{-0.0004}$ & $90^{+28}_{-30}$ & $0.11^{+0.2}_{-0.09}$ \\
J1632+3500 & $248.044^{+0.002}_{-0.002}$ & $35.01^{+0.001}_{-0.001}$ & $6^{+14}_{-17}$ & $0.4^{+0.16}_{-0.19}$ \\
J1723+3411 & $260.9008^{+0.0004}_{-0.0004}$ & $34.1992^{+0.0004}_{-0.0004}$ & $132^{+36}_{-27}$ & $0.16^{+0.13}_{-0.14}$ \\
J2243$-$0935 (E) & $340.8395^{+0.0005}_{-0.0005}$ & $-9.5954^{+0.0003}_{-0.0002}$ & $10^{+2}_{-2}$ & $0.42^{+0.04}_{-0.03}$ \\
J2243$-$0935 (W) & $340.8395^{+0.0005}_{-0.0005}$ & $-9.5954^{+0.0003}_{-0.0002}$ & $10^{+2}_{-2}$ & $0.42^{+0.04}_{-0.03}$ \\
\enddata
\end{deluxetable*}

\subsubsection{BCG-ICL Transition Radius Selection}\label{sec:trans_icl_bcg}

Identifying the BCG-ICL transition radius is non-trivial and subject to systematics that result from the data reduction methods used to produce final science images \citep{Zhang2019}. Favored methods include a surface brightness cut \citep[e.g.,][]{Furnell2021} or a graphical interpretation of the light profile of the BCG+ICL system in which authors identify the ICL by the point at which the profile plateaus \citep[e.g.,][]{Kluge2021}. \citet{Contini2022} found that the transition radius usually occurs at 60$\pm$40 kpc.

It is important that the ICL measurements are high signal-to-noise, such that we can accurately fit an ellipse to the lower isophote values that approach the background noise level. For this reason, we use a transition radius of 30 kpc as was used in \citet{Kluge2021} for measuring shapes and centroids of the BCG and ICL. This radius is approximately consistent with a by-eye inspection of the masked light profiles of the BCG+ICL systems of the cluster sample (see appendix \ref{appendix:bcg+icl}). It also provides us with a consistent way across the whole sample to measure the shape and centroid of the BCG and ICL in the respective regimes in which they dominate.

\subsection{Chandra Data and ICM Measurement}
\label{sec:icm}

In order to make measurements on the ICM, we process raw \emph{Chandra} ACIS-I/S X-ray event data using the publicly available \texttt{CIAO} tools for data reduction. The \emph{Chandra} data reduction process, which produced images for ICM science measurements, is detailed in \citet{Gassis2025}. The resulting processed ICM distribution was modeled using the \texttt{CIAO Sherpa} modeling functions, applying one or two elliptical Gaussian distributions to model the ICM while simultaneously fitting the background. In cases where two objects were fit to the ICM distribution, we take the radially larger component as the ICM considered in this paper. For our purposes, we are concerned with the large-scale ICM component that should be gravitationally bound to the cluster-scale main DM halo. Applying \texttt{Sherpa} modeling with a Gaussian input allows us to capture the large-scale ICM shape and centroid up to the point that it remains distinguishable from the background X-ray counts. We use the cash fitting statistic \citep{Cash1979} to derive the best-fit parameters and $1\sigma$ deviation for each object. We report the resulting best-fit models and associated parameters in Table \ref{tab:icm_meas}.

\subsection{Strong Lensing Models and CLM Measurement}
\label{sec:clm}

The strong lensing models for these galaxy clusters were derived using the publicly available \texttt{Lenstool} software \citep{Jullo2007}. This tool uses MCMC chains to minimize the scatter between the observed image plane and that predicted by the outputted \texttt{Lenstool} model. With these lensing models, we are able to describe the DM distribution of the cluster lens using a linear combination of pseudo-isothermal ellipsoidal mass distributions. The lensing models are described in \citet{Sharon2020,Sharon2022a,Sharon2022b}.

We derived the 2D mass projection of the cluster-scale DM halo by sampling the \texttt{Lenstool}-generated MCMC posterior distribution with the derived lensing input parameters. In the models, there is an obvious core cluster that dominates the distribution at large scales, the CLM. The CLM is what we would expect to align with the BCG, ICL, and ICM components. Each iteration of the MCMC chain outputs a set of parameters that describe the CLM. We select the parameters from the most likely image plane model (lowest $\chi^2$ model) to be the true values of the position angle, ellipticity, and centroid of the CLM. We estimate the $5^{\mathrm{th}}$ and $95^{\mathrm{th}}$ percentiles of each parameter’s MCMC posterior, corresponding to the 2$\sigma$ confidence interval, and define the lower uncertainty as the difference between the best-fit value and $5^{\mathrm{th}}$ percentile, and the upper uncertainty as the difference between $95^{\mathrm{th}}$ percentile and the best-fit value. We list the resulting elliptical properties of the CLM distributions of these clusters in Table \ref{tab:clm_meas}.

\begin{deluxetable*}{ccccc}
\centerwidetable
\setlength{\tabcolsep}{12pt}
\def\arraystretch{1.2}
\tablecaption{This table lists the measured elliptical properties of the Core Lensing Mass (CLM). The right ascension and declination are reported in degrees (J2000). We report the position angle in degrees, measured counterclockwise from West ($0^\circ$ corresponds to West). We define the reported ellipticity according to equation~\ref{eq:eps}.\label{tab:clm_meas}}
\tablehead{
\colhead{Name} & \colhead{$\alpha_{CLM}$} & \colhead{$\delta_{CLM}$} & \colhead{PA$_{CLM}$} & \colhead{$\emph{e}_{CLM}$}
}
\startdata
J0004$-$0103 & $1.216^{+0.001}_{-6e-05}$ & $-1.054^{+0.002}_{-0.0001}$ & $131^{+24}_{-35}$ & $0.61^{+0.05}_{-0.49}$ \\
J0108+0623 & $17.1752^{+0.0002}_{-0.0003}$ & $6.4120^{+0.0004}_{-0.0001}$ & $170^{+16}_{-15}$ & $0.24^{+0.39}_{-0.2}$ \\
J0146$-$0929 & $26.73348^{+7e-05}_{-5e-05}$ & $-9.49767^{+6e-05}_{-2e-05}$ & $156^{+2}_{-1}$ & $0.13^{+0.02}_{-0.02}$ \\
J0150+2725 & $27.504^{+0.0004}_{-0.001}$ & $27.4260^{+0.0006}_{-0.0003}$ & $49^{+5}_{-4}$ & $0.13^{+0.01}_{-0.02}$ \\
J0333$-$0651 & $53.2681^{+0.0005}_{-0.0002}$ & $-6.8567^{+0.0003}_{-0.0001}$ & $153^{+3}_{-3}$ & $0.56^{+0.07}_{-0.15}$ \\
J0851+3331 & $132.9121^{+0.0001}_{-0.0002}$ & $33.5190^{+0.0002}_{-0.0001}$ & $36^{+0.6}_{-2}$ & $0.23^{+0.09}_{-0.05}$ \\
J0915+3826 & $138.914^{+6e-06}_{-0.002}$ & $38.4492^{+0.0007}_{-8e-05}$ & $27.0^{+0.7}_{-0.8}$ & $0.55^{+0.09}_{-0.18}$ \\
J0928+2031(NW) & $142.0193^{+0.0004}_{-0.0002}$ & $20.5284^{+0.0002}_{-0.0004}$ & $48^{+4}_{-4}$ & $0.19^{+0.06}_{-0.04}$ \\
J0928+2031 (SE) & $142.0193^{+0.0004}_{-0.0002}$ & $20.5284^{+0.0002}_{-0.0004}$ & $48^{+4}_{-4}$ & $0.19^{+0.06}_{-0.04}$ \\
J0952+3434 & $148.1675^{+0.0008}_{-0.0003}$ & $34.5803^{+0.0003}_{-0.0007}$ & $40^{+2}_{-1}$ & $0.35^{+0.16}_{-0.11}$ \\
J0957+0509 & $149.413^{+0.0006}_{-0.001}$ & $5.1589^{+0.0001}_{-0.0005}$ & $160^{+2}_{-2}$ & $0.61^{+0.16}_{-0.05}$ \\
J1038+4849 & $159.67979^{+9e-05}_{-9e-05}$ & $48.82205^{+2e-05}_{-3e-05}$ & $34^{+1}_{-1}$ & $0.35^{+0.03}_{-0.04}$ \\
J1050+0017 & $162.66654^{+4e-05}_{-7e-05}$ & $0.2862^{+0.0002}_{-0.0002}$ & $65^{+4}_{-4}$ & $0.05^{+0.02}_{-0.01}$ \\
J1055+5547 & $163.7695^{+0.0001}_{-0.0002}$ & $55.8071^{+0.0006}_{-0.0004}$ & $95^{+4}_{-2}$ & $0.54^{+0.05}_{-0.08}$ \\
J1110+6459 & $167.5736^{+0.0001}_{-0.0001}$ & $64.99658^{+4e-05}_{-6e-05}$ & $-20^{+10}_{-3}$ & $0.60^{+0.1}_{-0.16}$ \\
J1115+1645 & $168.7689^{+0.0002}_{-0.0001}$ & $16.761^{+0.001}_{-0.0002}$ & $102^{+5}_{-4}$ & $0.41^{+0.09}_{-0.06}$ \\
J1138+2754 & $174.5374^{+0.0002}_{-5e-05}$ & $27.9081^{+0.0002}_{-0.0004}$ & $77.0^{+0.8}_{-0.9}$ & $0.59^{+0.06}_{-0.04}$ \\
J1152+3312 & $178.00137^{+5e-05}_{-8e-05}$ & $33.22863^{+4e-05}_{-8e-05}$ & $172^{+0.8}_{-1}$ & $0.114^{+0.002}_{-0.006}$ \\
J1152+0930 & $178.19617^{+9e-05}_{-3e-05}$ & $9.5056^{+9e-05}_{-0.0002}$ & $49.0^{+0.0}_{-0.6}$ & $0.14^{+0.05}_{-0.004}$ \\
J1156+1911 & $179.0227^{+0.0005}_{-0.0005}$ & $19.1866^{+0.0008}_{-0.0002}$ & $97^{+18}_{-14}$ & $0.14^{+0.51}_{-0.11}$ \\
J1207+5254 & $181.900^{+0.0005}_{-0.001}$ & $52.918^{+0.0006}_{-0.002}$ & $113^{+6}_{-9}$ & $0.25^{+0.4}_{-0.17}$ \\
J1209+2640 & $182.34878^{+4e-05}_{-4e-05}$ & $26.67940^{+4e-05}_{-4e-05}$ & $137^{+2}_{-3}$ & $0.21^{+0.01}_{-0.02}$ \\
J1226+2149 & $186.7124^{+0.0003}_{-0.0003}$ & $21.8316^{+0.0002}_{-0.0002}$ & $25^{+2}_{-5}$ & $0.15^{+0.05}_{-0.04}$ \\
J1226+2152 & $186.7154^{+0.0001}_{-6e-05}$ & $21.8733^{+3e-05}_{-0.0002}$ & $36^{+33}_{-6}$ & $0.03^{+0.01}_{-0.03}$ \\
J1329+2243 & $202.3935^{+0.0001}_{-9e-05}$ & $22.7228^{+0.0004}_{-0.0008}$ & $94^{+2}_{-2}$ & $0.19^{+0.07}_{-0.0001}$ \\
J1336$-$0331 & $204.0011^{+4e-05}_{-0.0004}$ & $-3.5255^{+0.0002}_{-3e-05}$ & $57^{+2}_{-16}$ & $0.03^{+0.07}_{-0.004}$ \\
J1343+4155 & $205.889^{+0.0008}_{-0.002}$ & $41.9184^{+0.0004}_{-0.0008}$ & $156^{+0.9}_{-3}$ & $0.24^{+0.39}_{-0.04}$ \\
J1420+3955 & $215.1677^{+0.0002}_{-0.0001}$ & $39.9201^{+9e-05}_{-0.0003}$ & $175^{+36}_{-14}$ & $0.02^{+0.1}_{-0.01}$ \\
J1429+1202 & $217.4782^{+0.0006}_{-0.0003}$ & $12.0433^{+0.0006}_{-0.0009}$ & $80^{+6}_{-7}$ & $0.48^{+0.17}_{-0.31}$ \\
J1439+1208 & $219.7912^{+0.0002}_{-0.0002}$ & $12.1403^{+0.0001}_{-0.0002}$ & $19^{+2}_{-1}$ & $0.09^{+0.04}_{-0.02}$ \\
J1456+5702 & $224.0036^{+0.0001}_{-7e-05}$ & $57.0388^{+0.0002}_{-0.0001}$ & $85^{+4}_{-7}$ & $0.15^{+0.11}_{-0.03}$ \\
J1522+2535 & $230.7193^{+0.0002}_{-0.0002}$ & $25.5914^{+0.0001}_{-0.0002}$ & $39^{+2}_{-1}$ & $0.07^{+0.04}_{-0.01}$ \\
J1531+3414 & $232.7939^{+0.0001}_{-7e-05}$ & $34.2401^{+9e-05}_{-0.0001}$ & $83^{+12}_{-12}$ & $0.12^{+0.04}_{-0.03}$ \\
PSZ1G311 & $237.5294^{+6e-05}_{-0.0001}$ & $-78.19163^{+9e-05}_{-8e-05}$ & $0^{+2}_{-0.4}$ & $0.23^{+0.01}_{-0.02}$ \\
J1604+2244 & $241.0415^{+0.0006}_{-0.0004}$ & $22.7368^{+0.0004}_{-0.0005}$ & $138^{+7}_{-2}$ & $0.58^{+0.08}_{-0.37}$ \\
J1621+0607 & $245.3850^{+0.0002}_{-5e-05}$ & $6.1225^{+0.0001}_{-0.0004}$ & $76^{+3}_{-2}$ & $0.16^{+0.03}_{-0.009}$ \\
J1632+3500 & $248.045^{+6e-05}_{-0.001}$ & $35.0090^{+0.0002}_{-6e-05}$ & $188^{+5}_{-3}$ & $0.23^{+0.33}_{-0.004}$ \\
J1723+3411 & $260.9003^{+0.0008}_{-0.0009}$ & $34.1996^{+0.0003}_{-7e-05}$ & $-3^{+3}_{-2}$ & $0.31^{+0.12}_{-0.07}$ \\
J2111$-$0114 & $317.8309^{+0.0007}_{-9e-05}$ & $-1.240^{+0.003}_{-0.0004}$ & $102^{+2}_{-1.0}$ & $0.21^{+0.29}_{-0.04}$ \\
J2243m0935 (W) & $340.8334^{+0.0006}_{-0.0003}$ & $-9.5907^{+0.0002}_{-0.0002}$ & $151^{+0.5}_{-17}$ & $0.33^{+0.005}_{-0.17}$ \\
J2243$-$0935 (E) & $340.8572^{+0.0003}_{-0.0009}$ & $-9.582^{+0.001}_{-0.001}$ & $108^{+0.8}_{-8}$ & $0.55^{+0.02}_{-0.08}$ \\
\enddata
\end{deluxetable*}

\section{Results}
\label{sec:results}

In Section~\ref{sec:dat-meas}, we describe the methods used to isolate the BCG, ICL, ICM, and CLM distributions and measure their elliptical properties. Using these ellipse-fitting techniques, we compare the shapes (position angles and ellipticities) and centroids of the various components to quantify the degree of alignment between the different cluster distributions. 

\begin{figure}
	\includegraphics[width=\columnwidth]{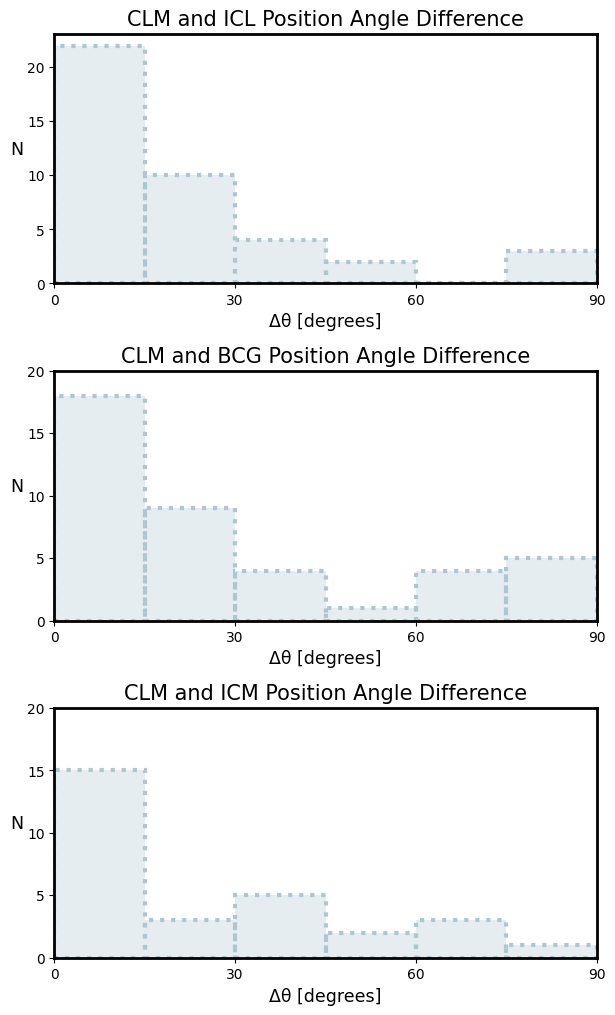}
    \caption{The difference in position angle of the major axis between the CLM and the three other distributions (from top to bottom ICL, BCG, and ICM).}
    \label{fig:pa_mass}
\end{figure}

\begin{figure}
	\includegraphics[width=\columnwidth]{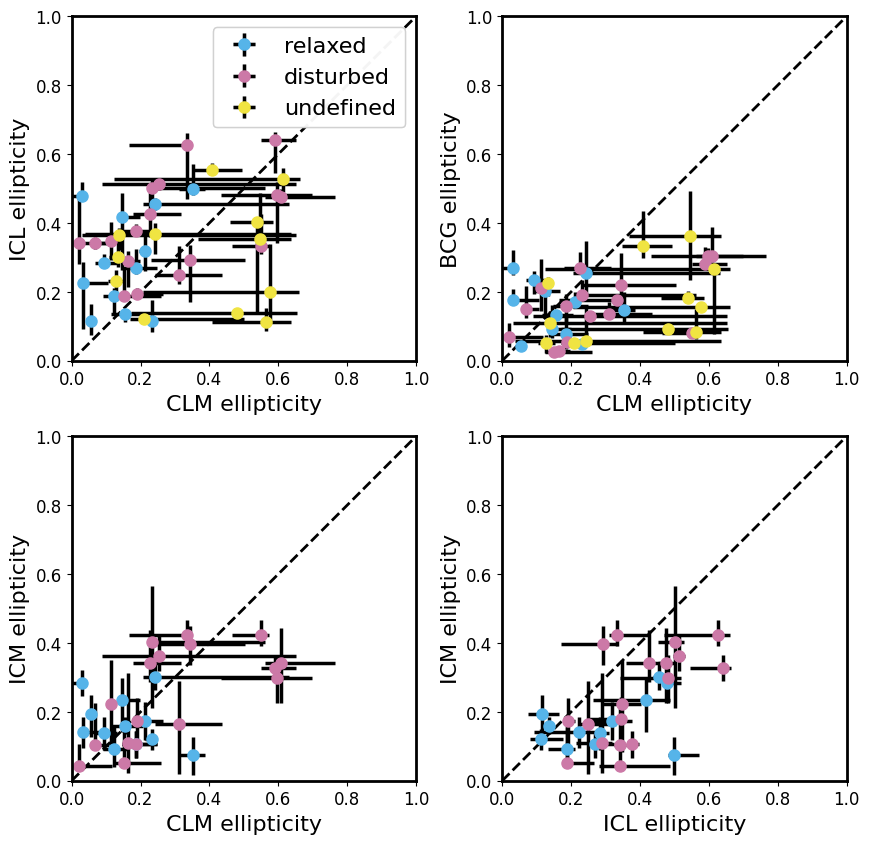} 
    \caption{Top-left: the ellipticity comparison between the CLM and ICL, top-right: the ellipticity comparison between the CLM and BCG, bottom-left: the ellipticity comparison between the CLM and ICM, bottom-right: the ellipticity comparison between the ICL and ICM. We illustrate the one-to-one line in black line. We differentiate clusters based on their dynamical state determination.}
    \label{fig:eps_mass}
\end{figure}

\begin{figure}
	\includegraphics[width=\columnwidth]{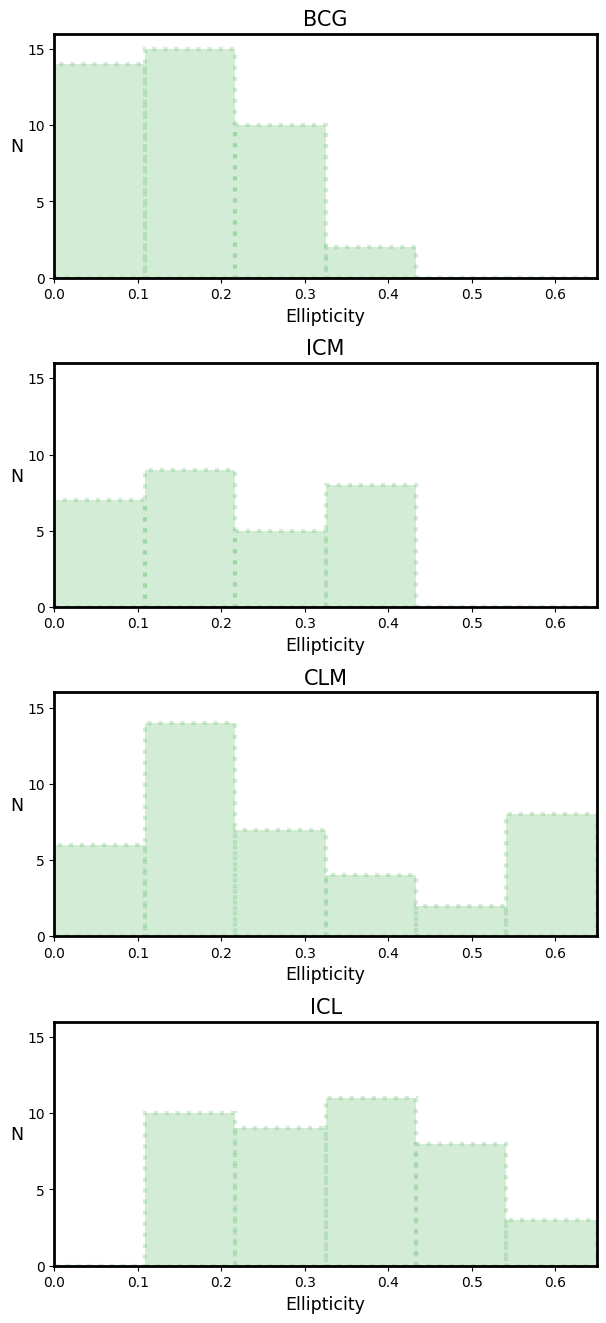}
    \caption{The ellipticity measurements ordered from most spherical (top) to most elliptical (bottom). In order (top-bottom), we display the BCG, ICM, CLM, and ICL ellipticity values.}
    \label{fig:eps_hist}
\end{figure}

\begin{figure*}
	\includegraphics[width=\textwidth]{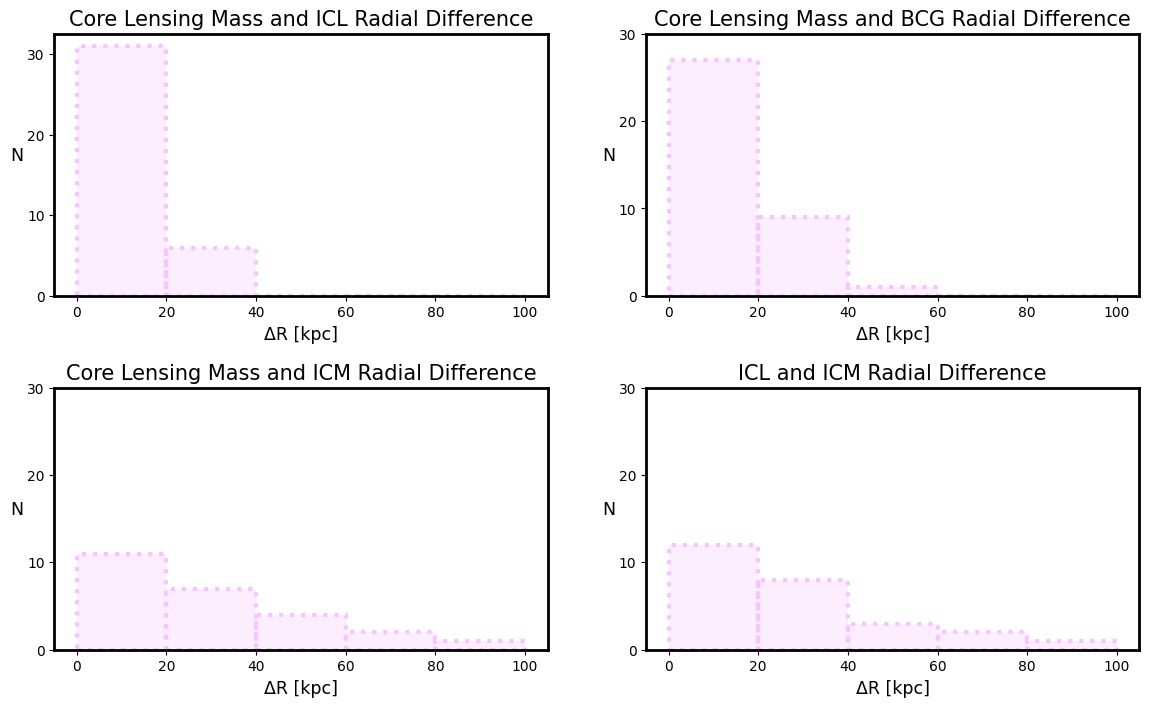}
    \caption{The comparison of centroid deviation in kpc of various components of the galaxy cluster systems in this sample. We compare the centroids of the ICL and CLM (top-left), the BCG and the CLM (top-right), the ICM and the CLM (bottom-left), and the ICM and ICL (bottom-right). For the purpose of visualization, we do not include double-cored systems.}
    \label{fig:cent_mass}
\end{figure*}

\subsection{Position Angles}
\label{sec:pa}

In Figure~\ref{fig:pa_mass}, we compare the measurements of the position angles of the BCG, ICL, and ICM with the position angle of the CLM. For all components, we measure the position angle in degrees counterclockwise from West ($0^\circ$ corresponds to West). We define the difference in position angle $\Delta$PA by equation~\ref{eq:pa}.

\begin{align}
\label{eq:pa}
\begin{split}
 \Delta\mathrm{PA}=|\mathrm{PA}_1-\mathrm{PA}_2| \;\;\;\;\;\;\;\;\;
 (|\mathrm{PA}_1-\mathrm{PA}_2|\le90^\circ)
\\
 \Delta\mathrm{PA}=180-|\mathrm{PA}_1-\mathrm{PA}_2| \;\;\;\;\;\;\;\;\;
 (|\mathrm{PA}_1-\mathrm{PA}_2|>90^\circ)
 \\
0^\circ \le \Delta\mathrm{PA}\le90^\circ \;\;\;\;\;\;\;\;\;\;\;\;\;\;\;\;\;\;\;\;\;\;\;\;\;\;\;\;\;\;\;\;\;\;\;\;\;\;\;\;\;
\end{split}
\end{align}

Generally, we measure small position angle differences, which implies that cluster orientation is consistent over a large spatial scale. Though there are variations in size across the sample, this generally implies that there is consistent alignment throughout the various components of the cluster and their respective spatial scales, from a few tens of kpc up to $\sim$1Mpc. This type of behavior is consistent with previous studies \citep[e.g.,][]{Donahue2016}.

However, we do measure a significant amount of large position angle differences which we classify as $\Delta\mathrm{PA}>30^{\circ}$. We find that the percentage of high position angle differences is $22\%^{+8}_{-5}$, $34\%^{+8}_{-6}$, and $38\%^{+9}_{-8}$ for CLM-ICL, CLM-BCG, and CLM-ICM comparisons, respectively. From here on, all percentage errors are the Bayesian binomial confidence intervals drawn from the $1\sigma$ confidence interval for the beta distribution \citep{Cameron2011}.

The relatively small number of deviations between the CLM and ICL suggest that the ICL may be a more viable proxy for the orientation of the DM halo distribution in many cases. This is likely because the ICL is dominantly bound to the DM halo and is not influenced by the cluster members \citep{Niederste-Ostholt2010,Hashimoto2014,Donahue2016}.

The BCG performs worse than the ICL, even though previous authors have theorized that mergers happen preferentially along the major axis of the CLM \citep{Ragone-Figueroa2020}. This implies that the preference for mergers along the axis is not a guaranty. An imbalance in the direction of mergers can leave an imprint on the BCG resulting in an orientation misaligned from the CLM. However, as we will describe in Sections~\ref{sec:eps} and~\ref{sec:phys-con}, we can explain many cases of BCG misalignments with respect to the CLM via BCG sphericalization. In any case, this misalignment is in consequence of the fact that the BCG is a dense, self-gravitating object that does not have the ability to ``flow out" into the shape of the DM halo as easily as the ICL and ICM.

The ICM also has a greater number of deviations than the ICL despite the fact that they are both subject to the gravitational force of the cluster itself as opposed to individual cluster members. The self-interacting nature of the ICM likely causes the more frequent misalignments, which can lead it to deviate from the orientation of the CLM. Also, the ICM's increased sensitivity to mergers will cause it to experience greater deviations generally.

\subsection{Ellipticities}
\label{sec:eps}

In Figure~\ref{fig:eps_mass}, we compare the ellipticity measurements of the different components of the clusters in this sample. We use the definition of ellipticity analogous to the flattening parameter \citep{Binney1998} defined by Equation~\ref{eq:eps} where $a$ and $b$ are the semi-major axis and the semi-minor axis, respectively.

\begin{equation}
e=1-\frac{b}{a}
\label{eq:eps}
\end{equation}

Note that \texttt{Lenstool} outputs ellipticity in the form of Equation~\ref{eq:eps_lens}. We transform the outputted CLM \texttt{Lenstool} ellipticity ($\epsilon$) into the ellipticity in the form of Equation~\ref{eq:eps} ($e$) by solving for the axis ratio and transforming it for accurate comparison to the other cluster components. 

\begin{equation}
\epsilon=\frac{a^2-b^2}{a^2+b^2}
\label{eq:eps_lens}
\end{equation}

We look for a distribution's bias to be more or less elliptical using the bi-weight location of the difference in ellipticities ($\Delta e_{avg}$) between the two components \citep{Beers1990}. It is worth noting that the CLM ellipticities are subject to larger uncertainties as a consequence of degeneracy in terms of a lensing profile's ability to reconstruct lensing features over a relatively large range of ellipticity values. This degeneracy and the related uncertainty become more prevalent in more elliptical DM halos. Also, we note that in Figure \ref{fig:eps_mass} we include the dynamical state determination described in detail in section \ref{sec:dyn-st} to highlight that typically relaxed clusters are less elliptical for all cluster components except the BCG.

From the top-left panel of Figure~\ref{fig:eps_mass}, we see that the ICL is not strongly biased to be more or less elliptical than the DM distribution. There is a slight preference for higher ellipticity values in the ICL which is emphasized by the bi-weight location of $\Delta e_{avg,ICL-CLM}=0.08(3)$. We would expect the ICL to trace the gravitational distribution of the cluster rather than the shape of the individual cluster member galaxies \citep[e.g.,][]{Montes2018}; however, we do not see a clear relation between these CLM and ICL ellipticities. The scatter in this relation may be due to the relatively large uncertainty in CLM ellipticities, the projection effects leading to perceived elongation/sphericalization of the ICL, and/or the difference in scale between the ICL and CLM (the detectable ICL contained to smaller radii).

In the top-right panel of Figure~\ref{fig:eps_mass}, we observe that the BCG tends to be noticeably more spherical than the CLM with $\Delta e_{avg,BCG-CLM}=-0.11(3)$. This is likely a result of dynamical friction effects driven by high stellar densities in the BCG \citep*{Arena2006}. Projection effects may also contribute to the perceived sphericalization. In any case, our results suggest that the BCG is a poor proxy for the underlying ellipticity of the cluster.

The bottom-left panel of Figure~\ref{fig:eps_mass} reveals that, in many cases, the ICM most closely matches the CLM ellipticity values with $\Delta e_{avg,ICM-CLM}=-0.007(0.03)$. There is still a small tendency for some clusters to have a more spherical ICM relative to the CLM. This could be a consequence of hydrodynamical sphericalization in the ICM \citep{Markevitch2007}.

In the bottom-right panel of Figure~\ref{fig:eps_mass}, we see that the ICL and ICM appear to be correlated with a bias for the ICM to be more spherical than the ICL ($\Delta e_{avg,ICL-ICM}=-0.14(2)$). This relationship is less scattered than the CLM comparisons, since we eliminate the scatter due to the large uncertainty in CLM ellipticity. This comparison emphasizes a clear difference between the hydrodynamically sphericalized ICM and more elongated ICL.

In Figure \ref{fig:eps_hist}, we display the distribution of the measured ellipticity values for each mass component in our sample. The ICL is the most elliptical, followed by the CLM and then the ICM, with the BCG being the most spherical. This is emphasized by the average ellipticity of each component measured using the bi-weight estimator: $e_{avg,ICL}=0.34(2)$, $e_{avg,CLM}=0.24(3)$, $e_{avg,ICM}=0.20(2)$, $e_{avg,BCG}=0.16(1)$. This corroborates the trends shown in Figure \ref{fig:eps_mass}, that the ICL is more elliptical than the CLM, the ICM is slightly more spherical than the CLM due to hydrodynamic effects, and the BCG is much more spherical than the CLM due to high stellar density.

\begin{deluxetable*}{ccccc}
\centerwidetable
\setlength{\tabcolsep}{10pt}
\def\arraystretch{1.2}
\tablecaption{Offsets in kpc between BCG, ICL, CLM, and ICM centroids.\label{tab:kpc-offset}}
\tablehead{
\colhead{Name} & \colhead{\emph{R}\textsubscript{BCG–CLM} [kpc]} & \colhead{\emph{R}\textsubscript{ICL–CLM} [kpc]} & \colhead{\emph{R}\textsubscript{ICM–CLM} [kpc]} & \colhead{\emph{R}\textsubscript{ICM–ICL} [kpc]}
}
\startdata
J0004$-$0103 & 4(37) & 24(34) & N/A & N/A \\
J0108+0623 & 4(12) & 4(12) & N/A & N/A \\
J0146$-$0929 & 6(2) & 1(2) & N/A & N/A \\
J0150+2725 & 16(18) & 16(19) & N/A & N/A \\
J0333$-$0651 & 37(17) & 34(17) & N/A & N/A \\
J0851+3331 & 12(6) & 7(7) & 69(32) & 68(39) \\
J0915+3826 & 25(25) & 5(23) & N/A & N/A \\
J0928+2031 (NW) & 11(7) & 7(7) & 22(10) & 27(8) \\
J0928+2031 (SE) & 138(7) & 157(8) & 22(10) & 156(9) \\
J0952+3434 & 15(17) & 16(17) & 57(28) & 43(22) \\
J0957+0509 & 1(27) & 4(35) & 19(42) & 23(25) \\
J1038+4849 & 26(2) & 5(3) & 8(5) & 10(7) \\
J1050+0017 & 21(8) & 18(9) & 9(19) & 13(16) \\
J1055+5547 & 14(20) & 20(21) & N/A & N/A \\
J1110+6459 & 2(3) & 3(14) & 24(19) & 21(27) \\
J1115+1645 & 13(9) & 13(17) & N/A & N/A \\
J1138+2754 & 9(12) & 10(13) & 48(19) & 58(15) \\
J1152+3312 & 11(2) & 5(5) & 73(30) & 78(33) \\
J1152+0930 & 44(5) & 39(7) & 58(27) & 22(24) \\
J1156+1911 & 6(24) & 1(23) & N/A & N/A \\
J1207+5254 & 24(34) & 10(28) & 94(35) & 92(17) \\
J1209+2640 & 6(2) & 16(4) & 38(10) & 23(12) \\
J1226+2149 & 13(11) & 10(11) & 13(14) & 6(7) \\
J1226+2152 & 9(5) & 9(6) & 5(9) & 12(8) \\
J1329+2243 & 36(23) & 39(22) & 7(21) & 39(21) \\
J1336$-$0331 & 9(4) & 6(8) & 11(10) & 9(12) \\
J1343+4155 & 32(33) & 23(32) & 33(33) & 11(14) \\
J1420+3955 & 21(9) & 16(9) & 17(13) & 12(15) \\
J1429+1202 & 5(30) & 3(23) & N/A & N/A \\
J1439+1208 & 10(8) & 15(8) & 26(12) & 19(10) \\
J1456+5702 & 4(7) & 7(8) & 16(16) & 12(21) \\
J1522+2535 & 16(8) & 18(9) & 43(21) & 51(21) \\
J1531+3414 & 5(3) & 1(7) & 4(4) & 5(6) \\
PSZ1G311 & 1(3) & 1(2) & 32(3) & 34(7) \\
J1604+2244 & 29(13) & 21(14) & N/A & N/A \\
J1621+0607 & 9(9) & 11(10) & 27(32) & 30(28) \\
J1632+3500 & 15(11) & 9(20) & 26(42) & 18(53) \\
J1723+3411 & 7(28) & 7(28) & 12(26) & 6(15) \\
J2111$-$0114 & 8(58) & 8(78) & N/A & N/A \\
J2243$-$0935 (W) & 75(14) & 60(16) & 158(18) & 150(16) \\
J2243$-$0935 (E) & 57(42) & 46(28) & 453(38) & 410(19) \\
\enddata
\end{deluxetable*}

\subsection{Centroids}
\label{sec:cent}

From Figure~\ref{fig:cent_mass}, we illustrate the centroid differences between the different components of the galaxy clusters in the sample when compared to one another (tabulated in Table \ref{tab:kpc-offset}). When calculating the difference in centroid, we use the projected radial distance in units of kpc (see equation~\ref{eq:proj}): 

\begin{align}
\mathrm{R}_{1-2}\,[\mathrm{kpc}] = d_A \Big[ &\left( (\alpha_{\mathrm{1}} - \alpha_{\mathrm{2}}) 
\cos(\delta_{\mathrm{2}}) \right)^2 \nonumber \\
&+ (\delta_{\mathrm{1}} - \delta_{\mathrm{2}})^2 \Big]^{1/2} 
\label{eq:proj}
\end{align}

In equation~\ref{eq:proj}, $\alpha$ and $\delta$ are the right ascension and declination in radians and $d_A$ is the angular diameter distance in units of kpc. An important note for Figure~\ref{fig:cent_mass} is that we excluded double-cored galaxy clusters (J2243$-$0935 and J0928+2031).

As expected, we measure small deviations in centroid for the ICL-CLM comparison as well as for the BCG-CLM comparison. This implies that generally the BCG lies close to the center of the DM gravitational potential and that the ICL builds up around the BCG. The ICL is slightly more aligned with the CLM distribution centroid than the BCG.

In $70\%^{+10}_{-17}$ of the systems where the BCG is offset from the CLM centroid by more than 20 kpc ($\Delta\mathrm{R}_{BCG-CLM} > 20$ kpc), the BCG and ICL centroids remain closely aligned, with $\Delta\mathrm{R}_{BCG-ICL} < 10$ kpc. In all but one of these cases, the ICL is more closely aligned with the CLM than the BCG is (i.e., $\Delta\mathrm{R}_{ICL-CLM} < \Delta\mathrm{R}_{BCG-CLM}$). In the one exception, the difference between the ICL and BCG offsets from the mass centroid is only $\sim$2 kpc.

In the remaining $30\%^{+17}_{-10}$ of cases where both the BCG is offset from the mass centroid by more than 20 kpc and the BCG–ICL offset exceeds 10 kpc, we find that the ICL remains closely aligned with the CLM centroid, with $\Delta\mathrm{R}_{ICL-CLM} < 10$ kpc.

From the combined analysis of $\Delta\mathrm{R}_{BCG-CLM}$, $\Delta\mathrm{R}_{ICL-CLM}$, and $\Delta\mathrm{R}_{BCG-ICL}$ we conclude that in a slight majority of cases, BCGs are displaced in tandem with their ICL counterpart, implying that the same mechanism causes them to be misaligned from the CLM. However, the ICL is not displaced as frequently or to as great of a magnitude as the BCG. This could be a consequence of the ICL's ability to relax back to the centroid of the CLM on a smaller timescale than the BCG, which has been theorized to exhibit residual wobbling even after the cluster has achieved relaxation \citep{Kim2017,Harvey2017}. This further implies that the ICL is less sensitive to dynamical disturbances than the BCG.

These ideas are furthered by the observation that the BCG can exhibit large centroid deviations in cases where the ICL remains aligned with the CLM distribution. Cases where both the BCG and ICL are miscentered with respect to the CLM may have undergone a dynamical disruption more recently than clusters with a displaced BCG but aligned ICL.

Our results show that the BCG is not always centroided with the CLM. The displacement of the BCG could introduce systematic uncertainty in the estimation of cluster masses via weak lensing \citep{Lange2018,Skibba2011}. These uncertainties would propagate into the uncertainties in cosmological parameters derived from the comparison of observed cluster mass abundances as a function of redshift to simulations with known cosmology and Cluster Mass Function (CMF) \citep[e.g.,][]{plank2016}.

Even after excluding double-cored systems, we still see that the ICM is often displaced to larger projected radii than the ICL or BCG when compared to the CLM centroid. We again emphasize that we use the large-scale ICM centroid derived from the ICM models described in section~\ref{sec:icm} as opposed to the X-ray peak position. The observed ICM offset is likely because the ICM “sloshes” as a consequence of major merger activity in the cluster’s past (\citealt*{Markevitch2001}; \citealt{Churazov2003,Johnson2012,Harvey2017}). We also see this ICM gas sloshing when we compare the ICM to the ICL distribution in the bottom right panel of Figure~\ref{fig:cent_mass}. This sloshing behavior persists even in cases where the BCG and ICL distributions have relaxed back to the CLM centroid since the ICM gas oscillations dissipate on a much longer time scale due to hydrodynamical processes that do not affect the stellar component counterparts \citep*{Markevitch2007}.

\begin{figure}
	\includegraphics[width=\columnwidth]{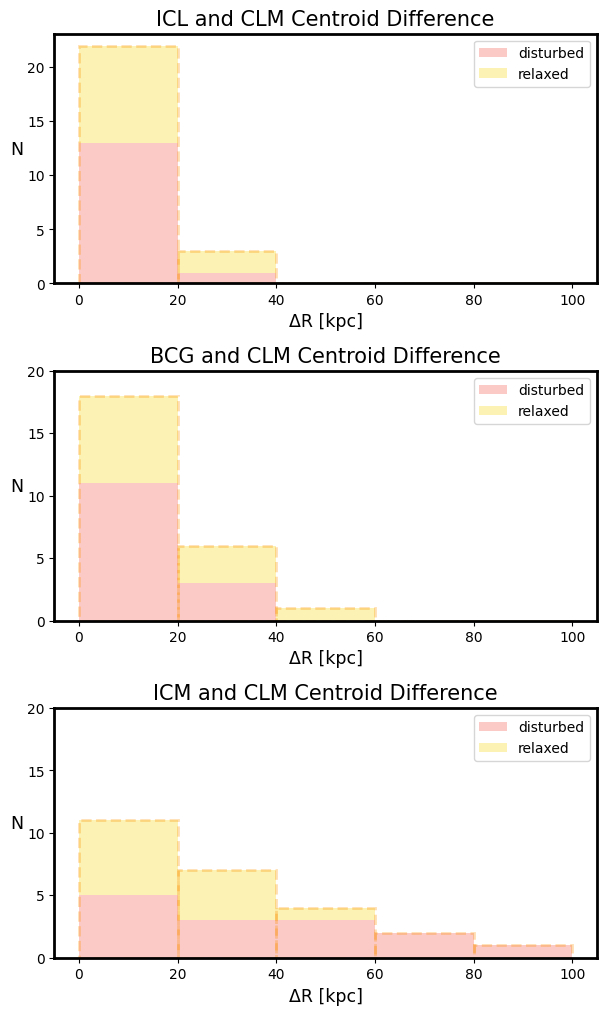}
    \caption{The comparison of centroid deviation in kpc of the ICL, BCG, and ICM (top to bottom) components of this galaxy cluster systems in this sample with respect to the CLM centroid. We only include systems where we have all four components measured simultaneously such that we are able to define the dynamical state using the ICM component. In these histograms, red indicates a disturbed dynamical state and yellow indicates a relaxed dynamical state. For the purpose of visualization, we do not include double-cored systems.}
    \label{fig:cent_sep_mass}
\end{figure}

\begin{figure}
\centering
	\includegraphics[width=.96\columnwidth]{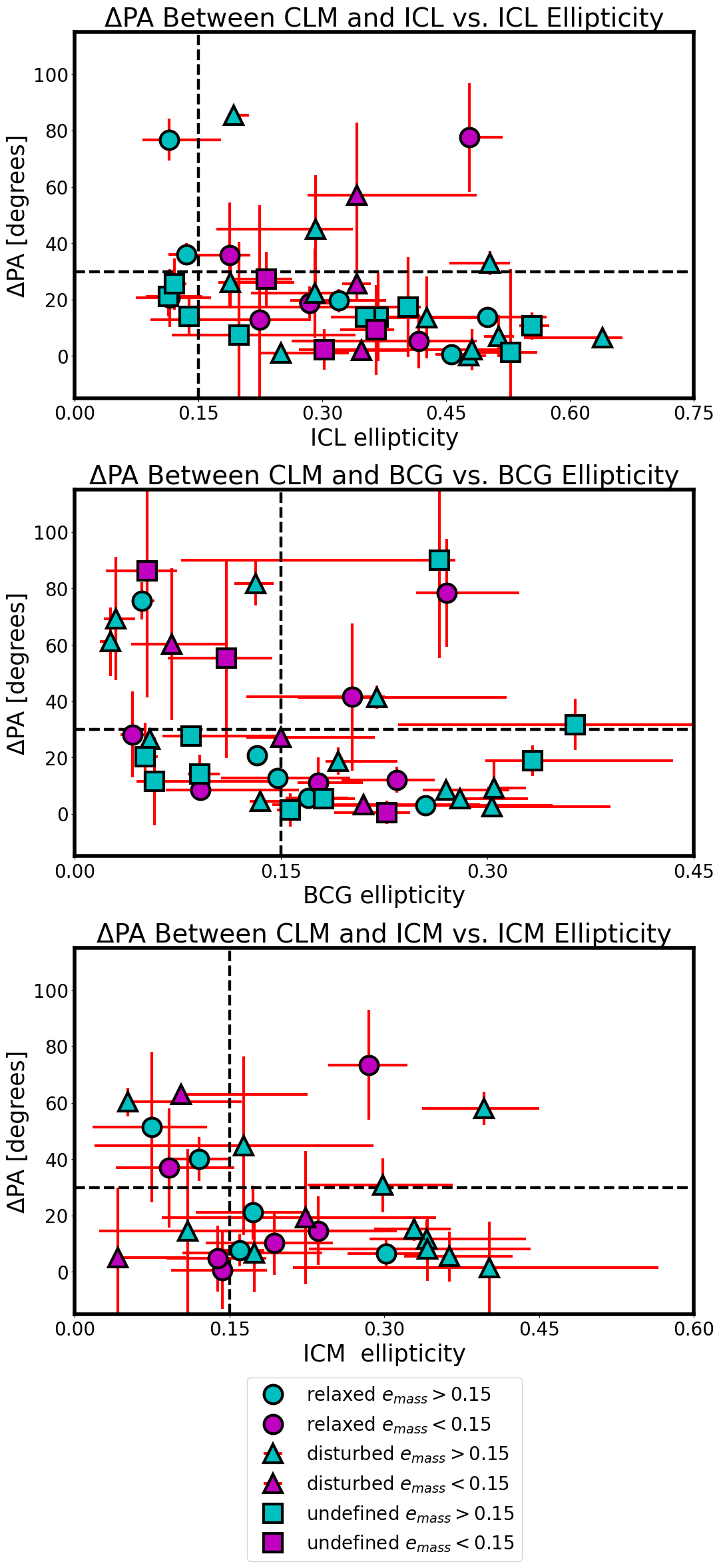}
    \caption{Top: the position angle difference between the CLM and the ICL distributions as a function of ICL ellipticity. Middle: the position angle difference between the CLM and the BCG distributions as a function of BCG ellipticity. Bottom: the position angle difference between the CLM and the ICM distributions as a function of ICM ellipticity. We distinguish points by their dynamical state: circles indicating relaxed clusters, triangles indicating disturbed clusters, and squares indicating an undefined dynamical state (no \emph{Chandra} Data). $\Delta$PA = $30^{\circ}$ and $e=0.15$ are highlighted by the dashed lines. We distinguish objects with more spherical DM halos by coloring objects with $e_{mass}>0.15$ cyan and object with $e_{mass}<0.15$ magenta.}
    \label{fig:mass_eps_cut}
\end{figure}

\section{Discussion}
\label{sec:disc}

From the results outlined in Section~\ref{sec:results}, we find general agreement with theoretical expectations: the BCG, ICL, and ICM are typically aligned with the CLM in regards to their position angle and centroid. We also observe a systematic trend in ellipticity, where the ICL tends to be the most elongated component, followed by the CLM and then the ICM, and the BCG generally appearing more spherical. Still, we observe occasional deviations from what is expected from $\Lambda$CDM predictions in cases of high position angle or centroid differences. We hope to understand these differences in the context of $\Lambda$CDM through multiple parameter analysis of the strong lensing cluster sample. 

We will do this by taking into consideration the dynamical state of the galaxy clusters for objects with available \emph{Chandra} X-ray data as well as the simultaneously considering all available measurements in our comparisons.

\subsection{Dynamical State}
\label{sec:dyn-st}

For objects with \emph{Chandra} X-ray data, we can use relaxation measurements to define the dynamical state of the cluster. This will allow us to determine how much the observed differences depend on whether the cluster is relaxed (non-merging/dynamically stable) or disturbed (merging/dynamically active).

\subsubsection{Dynamical State Measurement}
\label{sec:ds-meas}

In this Section, we will briefly introduce the parameter used to define cluster dynamical state. \citet{Gassis2025} contains a more detailed discussion of the morphological measurements used to classify the cluster dynamical state of the subsample of objects with \emph{Chandra} data.

We use the combined morphological parameter (\emph{M}) which incorporates measurements of the concentration parameter (\emph{c}), asymmetry parameter (\emph{A}), centroid shift parameter (\emph{log(w)}), and the X-ray–BCG centroid separation (\emph{D} [kpc]). Many of these parameters are computed within an aperture or a number of apertures related to $R_{500}$ which is the radius at which the mean density of a cluster is 500 times the critical density of the universe at the cluster's redshift. We take the equations used to calculate \emph{M}, \emph{c}, \emph{A}, \emph{log(w)}, and \emph{D} [kpc] and the corresponding values of these parameters from \citet{Gassis2025}.

\begin{figure}
\centering
	\includegraphics[width=\columnwidth]{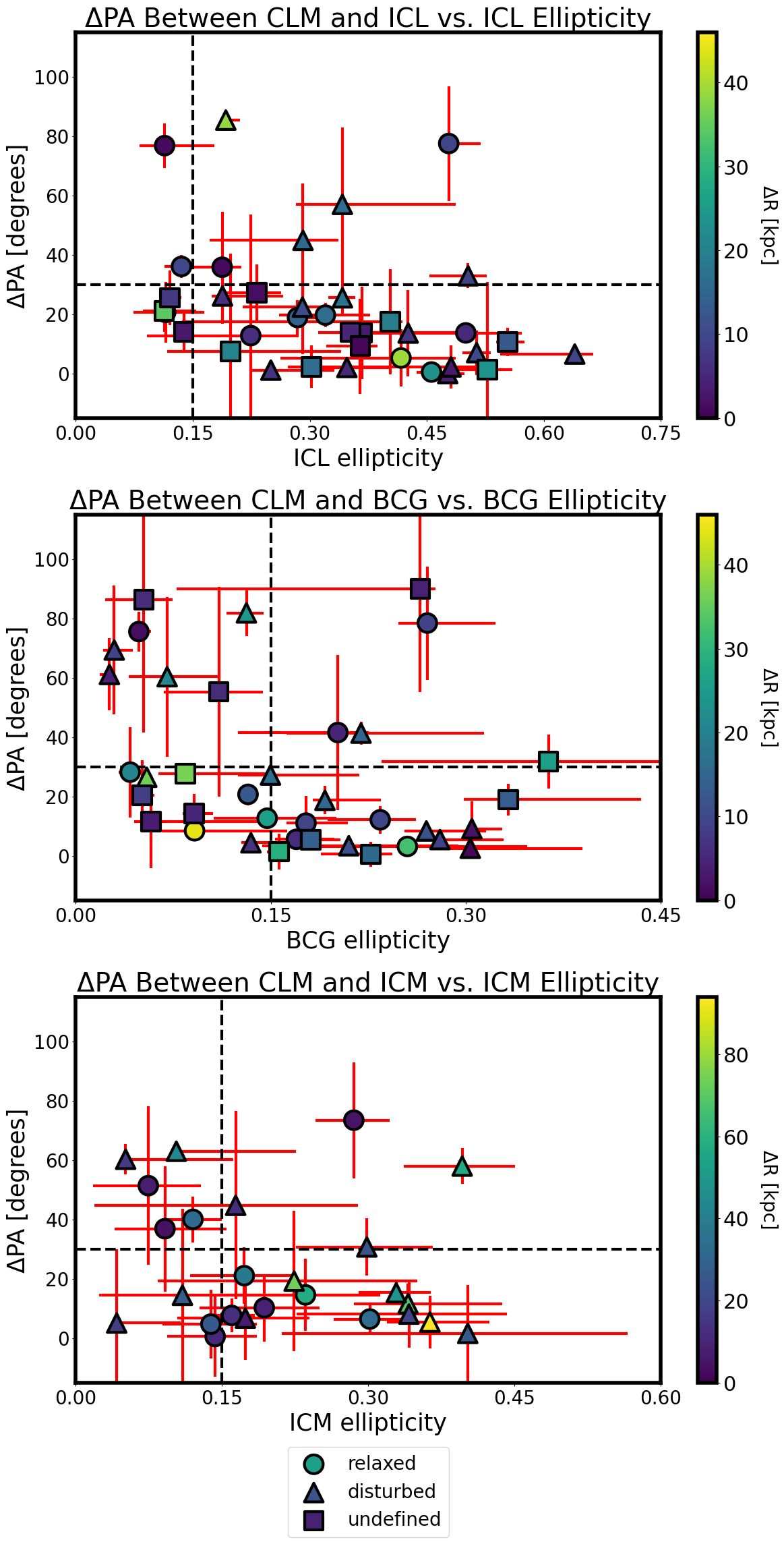}
    \caption{Top: the position angle difference between the CLM and the ICL distributions as a function of ICL ellipticity with radial offset of the centroids indicated by the color. Middle: the position angle difference between the CLM and the BCG distributions as a function of BCG ellipticity with radial offset of the centroids indicated by the color. Bottom: the position angle difference between the CLM and the ICM distributions as a function of ICM ellipticity with radial offset of the centroids indicated by the color. We distinguish points by their dynamical state: circles indicating relaxed clusters, triangles indicating disturbed clusters, and squares indicating an undefined dynamical state (no \emph{Chandra} Data). $\Delta$PA = $30^{\circ}$ and $e=0.15$ are highlighted by the dashed lines. For the purpose of visualization, we do not include double-cored systems.}
    \label{fig:kpc_col}
\end{figure}

\subsubsection{Dynamical State Analysis}
\label{sec:ds-an}

In Figure~\ref{fig:cent_sep_mass} we again analyze the centroid differences of the ICL, BCG, and ICM distributions with respect to the CLM distribution, this time incorporating the dynamical state of the cluster. In these histograms, we only include objects which have \emph{Chandra} X-ray data such that we were able to make the measurements described in~\ref{sec:ds-meas} to determine their dynamical state. J2243$-$0935 and J0928+2031 are excluded from this discussion due to their outlier nature as double-cored systems (see~\ref{sec:dc_sys}).

We do not detect a significant preference for cluster dynamical activity to cause misalignment between the BCG/ICL and the CLM. 
For the ICL and CLM centroid comparison, $18\%^{+16}_{-6}$ of the relaxed objects and $7\%^{+13}_{-2}$ of the disturbed objects have a centroid difference greater than 20 kpc. For the BCG-CLM Centroid comparison, $36\%^{+15}_{-11}$ of the relaxed objects and $21\%^{+14}_{-7}$ of the disturbed objects have a centroid difference greater than 20 kpc. The one relaxed cluster (J1152+0930) that has a large ICL-CLM centroid offset ($\Delta\mathrm{R}_{ICL-CLM} > 20$ kpc) \citet{Gassis2025} identifies as ``mixed more disturbed" with an $M=0.29(0.5)$ and also has a large BCG-CLM centroid offset ($\Delta\mathrm{R}_{BCG-CLM} > 20$ kpc). In any case, we conclude that dynamical disruptions typically do not result in large/persistent offsets in centroid between the ICL/BCG and the CLM.

In Section \ref{sec:cent}, we discussed how the BCG distribution is projected to larger displacements more often than the ICL. Still, the majority of both relaxed and disturbed clusters have small projected displacements, suggesting that though large deviations can occur, they resolve themselves over short timescales. The fact that we still observe large deviations for relaxed clusters favors the idea that the BCG can experience residual wobbling around an otherwise relaxed core \citep{Harvey2017,Harvey2019,Kim2017}.

Using this idea that the ICL may relax back to the shape of the CLM faster than the BCG and/or it is less displaced by dynamical activity in the cluster, we can conclude that the combined comparison of $\Delta\mathrm{R}_{BCG-CLM}$, $\Delta\mathrm{R}_{ICL-CLM}$, and $\Delta\mathrm{R}_{BCG-ICL}$ may give us a rough idea of the time since last major disruption event. Further work with larger cluster samples is needed to validate this claim.

For the ICM-CLM centroid comparison, $45\%^{+14}_{-13}$ of the relaxed objects and $64\%^{+10}_{-14}$ of the disturbed objects have a centroid difference ($\Delta\mathrm{R}_{ICM-CLM}$) greater than 20 kpc. This is consistent with the idea that the ICM is subject to persistent sloshing out of alignment with the cluster as a consequence of merger activity. Relaxed clusters can still show large centroid deviations between the ICM and CLM, but disturbed clusters have the ability to reach even greater deviations as a consequence of the ICM's sensitivity to merger activity.

In Table \ref{tab:KS_vals}, we further compare the measured parameters for the relaxed and disturbed subsamples using the average (median) value of each parameter for both subsamples. We use the median (with the error corresponding to the 1$\sigma$ confidence interval) as opposed to the bi-weight location since our relaxed subsample is composed of 11 clusters and our disturbed subsample is composed of 14 clusters, both of which are below the $N=15$ threshold suggested for use of the bi-weight estimator \citep{Beers1990}.

In the first 3 rows, we look at $\mathrm{R}_{1-2}$, $\Delta$PA, and $|\Delta e|$ for the ICL, BCG, and ICM compared to the CLM and investigate how those parameters differ for the relaxed and disturbed subsamples.

As previously discussed, the difference in centroids between the ICM and CLM is the only centroid component comparison that shows a noticeable variation between the relaxed and disturbed subsamples. The more disturbed clusters have larger centroid offsets on average. This is intuitive since the ICM remains displaced on long timescales due to merger activity. The BCG shows a slight trend in $\Delta$PA with respect to the CLM between the relaxed and disturbed systems. The more disturbed clusters have larger $\Delta$PA values on average. In addition, for $|\Delta e|$, we observe a modest difference between the relaxed and disturbed subsamples for the BCG and CLM comparisons, with more disturbed clusters having larger values of $|\Delta e|$ on average. The trends in $\Delta$PA and $|\Delta e|$ may be consequences of the CLM being elongated by merger activity in the direction of the merger and the BCG shape being less sensitive to cluster-scale events. We do not see any other significant differences in component comparisons between the relaxed and disturbed subsamples.

In addition to the component comparisons, in the last row of Table \ref{tab:KS_vals}, we also investigate the difference in the median ellipticity values for the relaxed and disturbed subsamples for each component. We see that the ICM has the largest difference in median values between the relaxed and disturbed subsamples, with the median ellipticity of the disturbed subsample being noticeably larger. This is intuitive since the large-scale ICM ellipticity has been used as a dynamical state proxy \citep[e.g.,][]{yuan2020,Yuan2023}. Though the BCG median ellipticity values are the same for the relaxed and disturbed subsamples (impacted more by its self-gravitation than cluster-scale gravitational effects), the ICL and CLM also have a detectable difference in median ellipticity between the relaxed and disturbed subsets, both having larger median ellipticity values for the disturbed subsample. This is reasonable because these components should be sensitive to cluster-scale disruptions, extending along the direction of merger events.

\begin{deluxetable}{c|ccc}
\setlength{\tabcolsep}{6pt}
\def\arraystretch{1.3}

\tablecaption{Comparison of the relaxed and distrurbed distributions of $\mathrm{R}{1-2}$, $\Delta$PA, $|\Delta e|$, and $e$, along with the average parameter values for the relaxed and disturbed subsamples. Average values of $\mathrm{R}{1-2}$, $\Delta$PA, and $|\Delta e|$ are reported for the ICL–CLM, BCG–CLM, and ICM–CLM comparisons, while $e$ values are provided for the ICL, BCG, ICM, and CLM individually. The reported averages correspond to the median, with uncertainties representing the 1$\sigma$ confidence interval around the median.
\label{tab:KS_vals}}
\tablehead{
\colhead{Parameter} & \colhead{Component(s)} & \colhead{Relaxed} & \colhead{Disturbed} \\
\colhead{} & \colhead{} & \colhead{Average} & \colhead{Average}}

\startdata
{} & ICL/CLM  & $10^{+10}_{-6}$ & $9^{+7}_{-4}$ \\
{$\mathrm{R}_{1-2}$} & BCG and CLM  & $10^{+19}_{-4}$ & $11^{+10}_{-7}$ \\
{} & ICM/CLM & $13^{+22}_{-6}$ & $26^{+42}_{-10}$ \\
\hline
{} & ICL/CLM & $20^{+33}_{-10}$ & $18^{+26}_{-16}$ \\
{$\Delta$PA} & BCG/CLM  & $13^{+43}_{-5}$ & $23^{+38}_{-18}$ \\
{} & ICM/CLM & $15^{+30}_{-9}$ & $15^{+42}_{-9}$ \\
\hline
{} & ICL/CLM & $0.15^{+0.09}_{-0.08}$ & $0.13^{+0.14}_{-0.08}$ \\
{$|\Delta e|$} & BCG/CLM & $0.08^{+0.12}_{-0.06}$ & $0.13^{+0.16}_{-0.07}$ \\
{} & ICM/CLM & $0.09^{+0.1}_{-0.05}$ & $0.11^{+0.15}_{-0.07}$ \\
\hline
{} & ICL & $0.28^{+0.18}_{-0.16}$ & $0.34^{+0.16}_{-0.09}$ \\
{$e$} & BCG & $0.17^{+0.07}_{-0.1}$ & $0.17^{+0.11}_{-0.12}$ \\
{} & ICM & $0.16^{+0.1}_{-0.05}$ & $0.26^{+0.1}_{-0.16}$ \\
{} & CLM & $0.14^{+0.09}_{-0.1}$ & $0.23^{+0.34}_{-0.11}$ \\
\enddata
\tablecomments{To maintain a homogeneous sample, we exclude data points associated to doubly-cored systems due their unique structural properties.}
\end{deluxetable}

\subsection{Multiple Parameter Analysis}
\label{sec:phys-con}

In Figure~\ref{fig:mass_eps_cut}, we combine the measurements made on each of the components of the galaxy cluster in reference to the measurements made on the CLM. Specifically, we plot the difference in position angle between each component and the CLM as a function of the component’s ellipticity. For consistency, we set boundaries of high position angle and low ellipticity (highly spherical) at $\Delta$PA = $30^{\circ}$ and $e=0.15$ respectively, for all components of the clusters. We use the same low ellipticity boundary of $e=0.15$ for the CLM component with magenta points being objects with more spherical ($e<0.15$) CLM distributions.

Figure~\ref{fig:kpc_col} is the same as Figure~\ref{fig:mass_eps_cut}, except we change the colors of the markers in order to show the effect displacement in centroid with respect to the CLM can have on frequency and magnitude of position angle deviations. As such, we remove the data points corresponding to J2243$-$0935 and J0928+2031 since at least one of their optical cores has a very large radial offset in centroid space compared to the rest of the sample. 

In these graphs, we continue to distinguish clusters based on their dynamical state by changing the shape of the plot marker based on whether the object is relaxed or disturbed. Objects with no X-ray data or a non-detection are labeled as undefined. The majority of this discussion will focus on the $\Delta$PA $>30^{\circ}$ and $e>0.15$ quadrant of the graph. 

In the upper panel of Figures~\ref{fig:mass_eps_cut} and~\ref{fig:kpc_col}, we analyze the ICL distribution. Generally, the ICL ellipticity is higher, which implies more meaningful measurements of the position angle for the ICL distribution. Still, there is a case of high position angle difference between the ICL and CLM as a consequence of low ICL ellipticity. Aside from that, we see good alignment with the CLM position angle measurement except for a few cases. Four of the six objects in the upper right quadrant of~\ref{fig:mass_eps_cut} are disturbed. When referencing the color bar in Figure~\ref{fig:kpc_col}, we also notice that the disturbed clusters in this quadrant are also moderately to extremely displaced, which signifies that dynamical disruptions can cause deviations in both centroid and position angle space simultaneously. When looking at Figure~\ref{fig:mass_eps_cut}, we see that the two relaxed objects that fall into this quadrant have CLM ellipticities less than 0.15, implying that the CLM distribution may be too spherical to give a meaningful position angle measurement for this cluster.

The middle panels of Figures~\ref{fig:mass_eps_cut} and~\ref{fig:kpc_col} compare the BCG to the CLM. Generally, we actualize the expected result from $\Lambda$CDM: for large differences in BCG and CLM position angles, we typically measure low BCG ellipticity as well. This likely reflects the fact that dynamical friction in BCGs can sphericalize their stellar mass distributions leading to less meaningful measurements of position angle in many cases which may explain why the BCG has higher position angle offsets than the other components with respect to the CLM. Position angles measured when ellipticities approach zero are not very meaningful. We observe that when BCGs become nearly spherical, with ellipticity $e_{BCG} < 0.15$, the corresponding $\Delta\mathrm{PA}_{BCG-CLM}$ tends to increase. In a few cases, we observe larger position angle differences despite a moderately elliptical BCG shape. The disturbed object in this quadrant also has moderately displaced centroids. This again suggests that the stellar components can be displaced in both centroid and position angle space simultaneously by dynamical disruptions. As was also the case for the ICL, the relaxed clusters in this quadrant have spherical CLM distributions.

For the lower panels, we realize the same L-shaped behavior as we did for the BCG distribution for the ICM-CLM position angle difference as a function of ICM ellipticity. Many objects with high position angle differences also have low ellipticities. Even though we typically see alignment for well-defined ICM ellipticities, there are some cases of misalignments for moderate ellipticity values. Most objects in this quadrant are disturbed. In Figure \ref{fig:kpc_col}, we see that two of the three disturbed clusters in the upper right quadrant have moderately/highly offset centroids. Also, in Figure \ref{fig:mass_eps_cut}, we see the relaxed cluster in this quadrant has a spherical CLM distribution. This behavior is consistent with the concept that the ICM is able to ``slosh" out of major axis alignment and/or centroid space due to merger events in the cluster's past.

\section{Conclusion}
\label{sec:con}

Generally, we find that strong lensing galaxy clusters show good alignment with $\Lambda$CDM predictions. In most cases, all cluster components align with the CLM, which we adopt as the reference for the true orientation, shape, and centroid of the galaxy cluster and its corresponding DM distribution. Usually, we can explain cases of misalignment can in the context of $\Lambda$CDM due to physical phenomena such as sphericalization of the BCG due to dynamical friction, sphericalization of the ICM as a result of hydrodynamical effects, ICM gas sloshing persistent from previous galaxy mergers, disturbed state of the galaxy cluster due to recent merger disruption, etc. Astrophysical phenomena like these may be able to explain some of the deviations from cluster component alignment found in former studies \citep[e.g.,][and the references therein]{Bosch2005,Harvey2017,Kim2017,Ragone-Figueroa2020,Contini2022}. The major conclusions of this paper are as follows:

\begin{itemize}
    \item{\emph{Position Angle Alignments of the Mass Components}}
    \begin{itemize}
        \item We find that orientation persists over a range of spatial scales (from a few tens of kpc up to $\sim$1Mpc). Components are generally well aligned in regards to their position angle orientation over a large spatial scale. This implies their formation histories are well-connected and interdependent as expected.
        \item We find that the ICL is best aligned with the DM halo as expected since physically the ICL stars behave much like the collisionless particles that make up CLM. Though the ICL can be displaced in position angle space due to merger activity, the ICL recovers from the disruptions most rapidly and in consequence maintains orientation with the cluster generally.
        \item When the BCG shows misalignment in position angle space with respect to the CLM, it is typically either a consequence of having a sphericalized distribution or being dynamically disrupted.
        \item Position angle deviations in the ICM are likely a consequence of the ICM being ``sloshed" out of position angle space due to dynamical disruptions or due to less meaningful position angle measurements as a consequence of hydrodynamical sphericalization.
    \end{itemize}

    \item{\emph{Elliptical Shape Trends of the Mass Components}}
    \begin{itemize}
        \item The ICL is the most elliptical mass component of the cluster followed by the CLM, ICM, and BCG.
        \item The BCG is more spherical than the cluster, which we interpret to be a consequence of dynamical friction due to high stellar densities in the BCG.
        \item The ICM aligns with the elliptical shape of the cluster in many cases. There are instances of ICM being more spherical likely due to hydrodynamical effects that influence the ICM shape.
    \end{itemize}

    \item{\emph{Centroid Alignments of the Mass Components}}
    \begin{itemize}
        \item The ICL and BCG centroid well with the CLM and with each other in most cases.
        \item The BCG may experience residual wobbling that continues for a longer period of time than it takes the ICL to relax back to the gravitational center of the cluster and as such the ICL is better centroided with the CLM.
        \item Our results indicate that displacements between the BCG and ICL components (and CLM when available) are sensitive to the time since last major merger event.
        \item The ICM is usually displaced to larger projected radii than the stellar components as a consequence of sensitivity to merger activity and persistent ``sloshing." \citep{Markevitch2001}
    \end{itemize}

    \item{\emph{Implications for Cluster Mass Tracing and Dynamical State}}
    \begin{itemize}
        \item Incorporating all of the results of this paper, we find that the ICL is most aligned with the CLM and, therefore, may be the optimal observable mass tracer of the cluster's underlying DM halo in absence of lensing information.
        \item The BCG exhibits noneligible deviations from the CLM centroid, which could introduce systematic uncertainties in studies that assume the BCG is the center of the cluster.
        \item Our results suggest that different cluster mass components are sensitive to merger activity on different timescales, with the ICL most likely to recover its stability in tandem with the CLM, followed by the BCG, and the ICM taking the longest to stabilize. 
    \end{itemize}

\end{itemize}

\vspace{1pt}

\section*{Acknowledgements}

R.G. would like to thank the listed coauthors for their support during this project and helpful input when writing the manuscript. This work was supported by grants associated with \emph{HST} programs GO-13003, GO-15377, and GO-15378; all of which were awarded by STScI under NASA contract NAS 5-26555. Support for this work was also provided by the National Aeronautics and Space Administration through \emph{Chandra} Award Number GO8-19111A issued by the \emph{Chandra} X-ray Center, operated by the Smithsonian Astrophysical Observatory for and on behalf of the National Aeronautics Space Administration under contract NAS8-03060. Additional support was provided by the Graduate College Dean's Dissertation Completion Fellowship awarded by the University of Cincinnati. This research has made use of data obtained from the \emph{Chandra} Data Archive provided by the \emph{Chandra} X-ray Center (CXC). The data set constituting the \emph{Chandra} Strong Lens Sample,  obtained by the \emph{Chandra} X-ray Observatory, is hosted at the \emph{Chandra} Data Collection (CDC) 434~\dataset[doi:10.25574/cdc.434]{https://doi.org/10.25574/cdc.434}. This research utilizes strong lensing models published in \citet{Sharon2020,Sharon2022a,Sharon2022b}. The reduced \emph{HST} images and lens models for the original SGAS-HST program are hosted at~\dataset[doi:10.17909/t9-cqtj-y020], on the Multimission Archive at the Space Telescope Science Institute (MAST) as detailed in \citet{Sharon2020}. The reduced \emph{HST} images and lens models for J1226+2152 and J1226+2149 are hosted at~\dataset[doi:10.17909/t9-cqtj-y020] on MAST as detailed in \citet{Sharon2022b}. \emph{HST} data for PSZ1G311 and J1429+1202 associated to GO-15377 and GO-15378 are hosted at~\dataset[doi:10.17909/jm83-gd60] on MAST.

\facilities{CXO (ACIS-I, ACIS-S), HST (WFC3), Sloan.}

\software{This research made use of pandas \citep{McKinney_2010, McKinney_2011}; Sherpa \citep{2001SPIE.4477...76F}; Astropy, a community-developed core Python package for Astronomy \citep{2018AJ....156..123A, 2013A&A...558A..33A}; SciPy \citep{Virtanen_2020}; matplotlib, a Python library for publication quality graphics \citep{Hunter:2007}; ds9, a tool for data visualization supported by the \emph{Chandra} X-ray Science Center (CXC) and the High Energy Astrophysics Science Archive Center (HEASARC) with support from the JWST Mission office at the Space Telescope Science Institute for 3D visualization; NumPy \citep{harris2020array}; the CIAO X-ray Analysis Software \citep{2006SPIE.6270E..1VF}; Astropy, a community-developed core Python package for Astronomy \citep{2018AJ....156..123A, 2013A&A...558A..33A}; Photutils \citep{Bradley2022}; pyregion, developed by Jae-Joon Lee (\url{https://github.com/leejjoon}); lsq-ellipse \citep{Hammel2020}; and GALFIT, a galaxy modeling algorithm software \citep{Peng2002}.
}

\bibliography{draft_4}{}
\bibliographystyle{aasjournal}

\begin{appendix}
\section{The BCG and ICL} \label{appendix:bcg+icl}
\begin{figure*}
	\includegraphics[width=\textwidth]{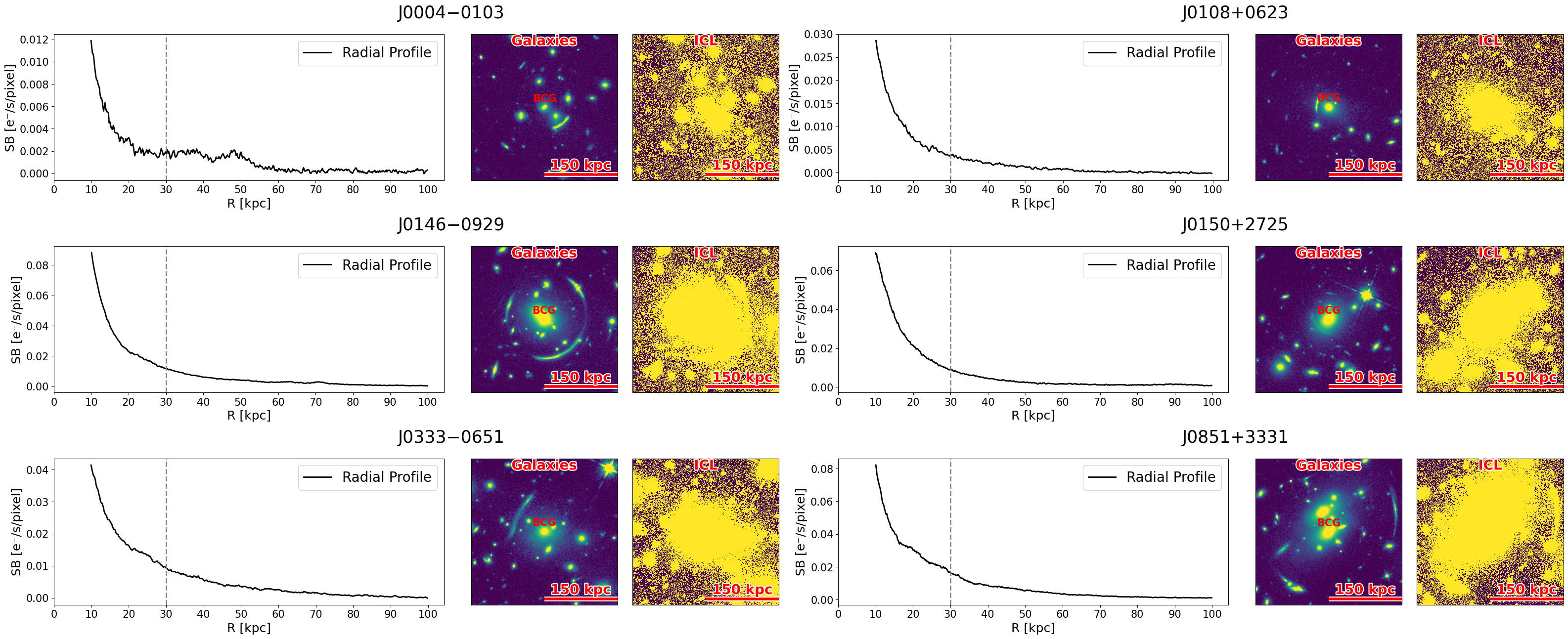}
    \caption{Rogues gallery, page 1 of 3. Here we display the BCG+ICL systems for each of our clusters. In each panel, the left plot shows the radial surface brightness profile of the BCG+ICL system, measured on the masked \emph{HST} WFC3 imaging data. The dashed gray line marks the 30 kpc transition radius used in this work, following \citet{Kluge2021}. The inner 10 kpc is excluded, as the surface brightness (SB) in this region is extremely high relative to the more diffuse light in the BCG outskirts and the ICL. The middle plot displays the full field, including the BCG, satellite galaxies, and any interloping sources. The right plot highlights the ICL.}
\label{fig:rogues1}
\end{figure*}

\begin{figure*}
	\includegraphics[width=\textwidth]{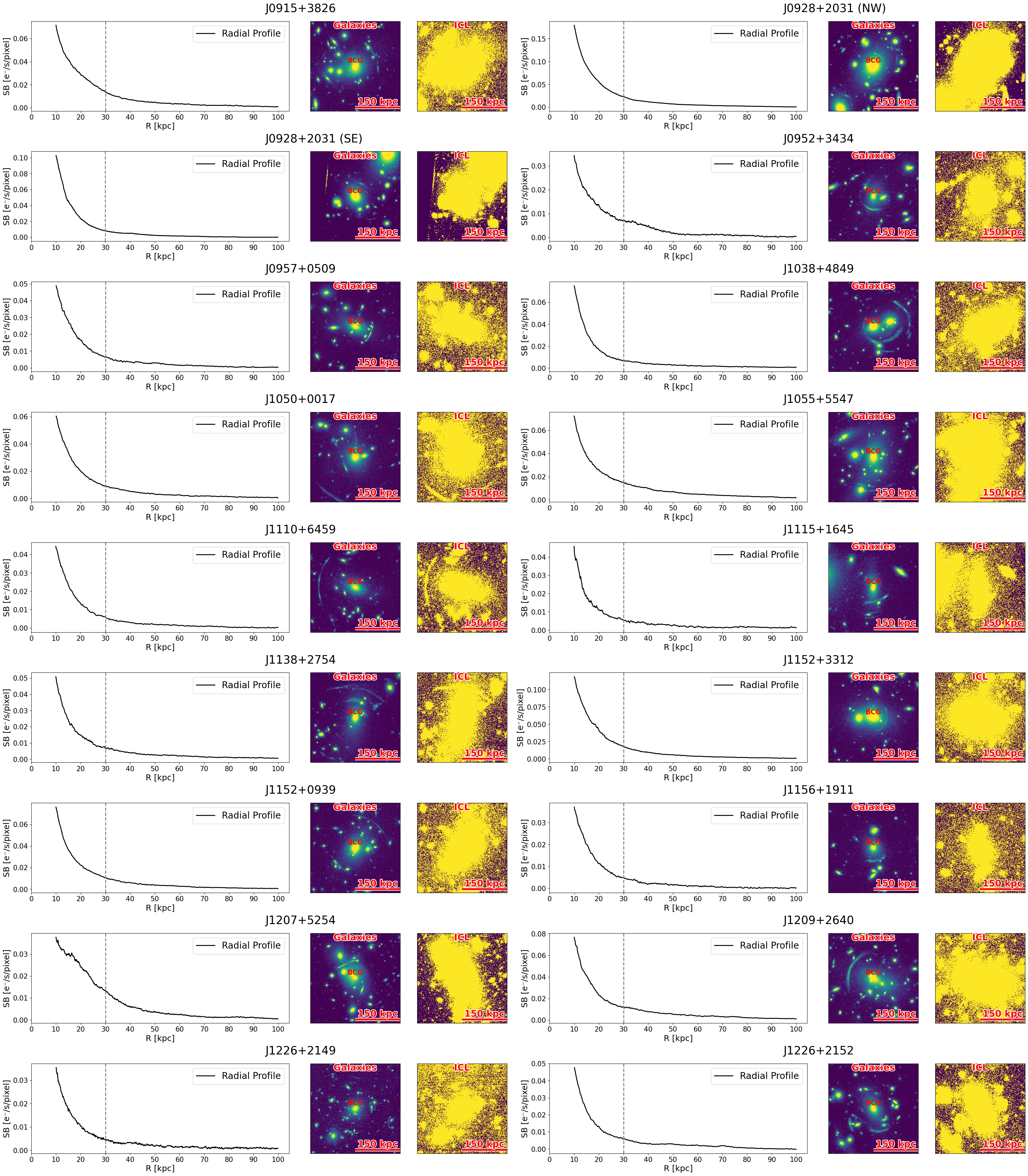}
    \caption{Rogues gallery, page 2 of 3. See figure \ref{fig:rogues1} for further details.}
\label{fig:rogues2}
\end{figure*}

\begin{figure*}
	\includegraphics[width=\textwidth]{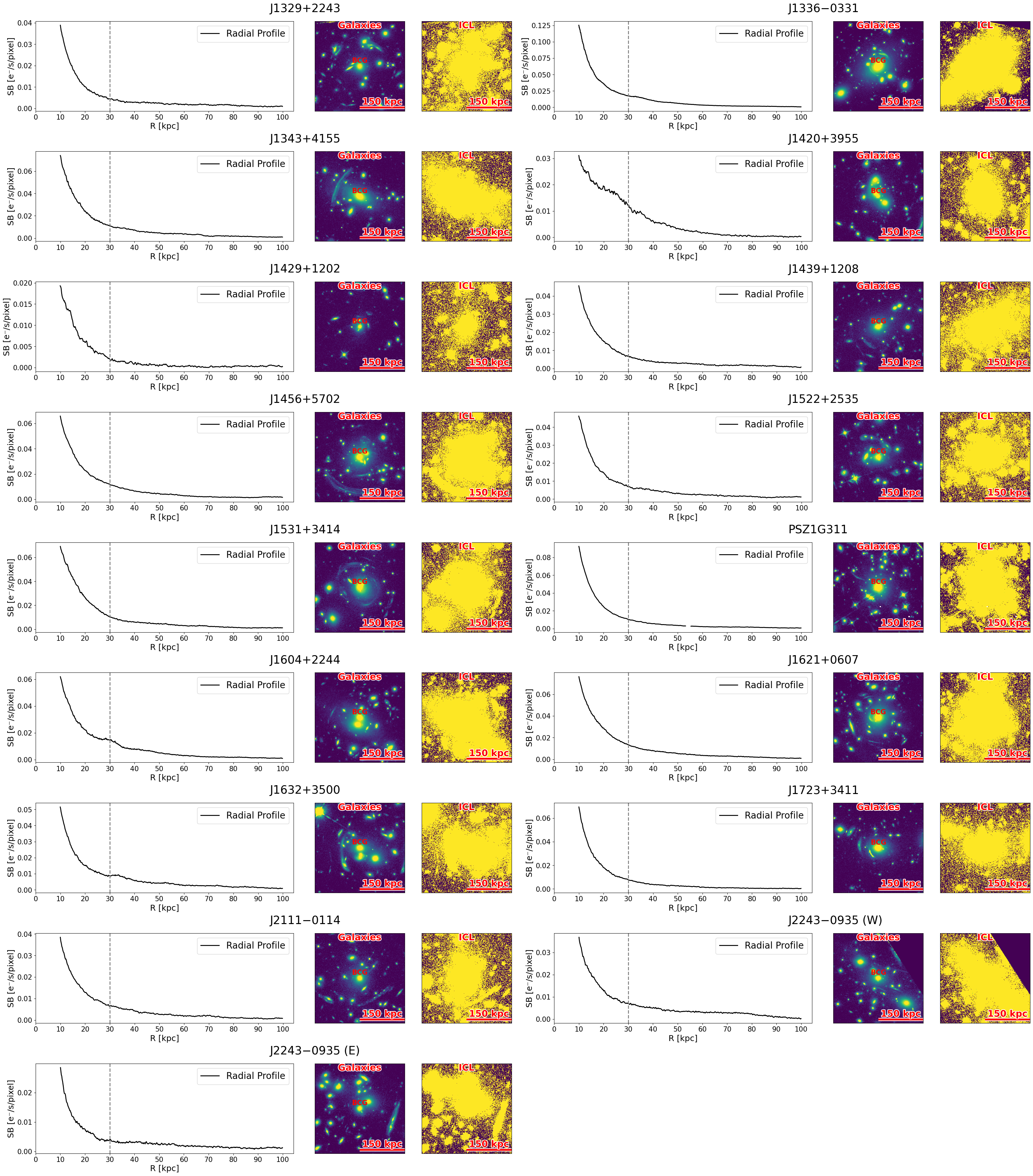}
    \caption{Rogues gallery, page 3 of 3. See figure \ref{fig:rogues1} for further details.}
\label{fig:rogues3}
\end{figure*}

As detailed in Sections \ref{sec:dat-meas} and \ref{sec:hst}, we perform measurements of the BCG and ICL on \emph{HST} WFC3 imaging data. The difference between these two entities is unambiguous in a theoretical sense: the BCG is all the stellar light subject to the gravitational influence of the BCG itself, whereas the ICL has been stripped from its original host galaxy and is now subject to the gravitational influence of the cluster as opposed to the BCG or any other cluster member. However, their formation histories are intertwined, since the mechanisms that dominantly contribute to ICL formation are concentrated around the gravitational center of the cluster wherein the BCG resides. As such, differentiating between the two entities presents an observational challenge. For the science goals of this project, we necessarily need a uniform method to differentiate between the BCG and ICL to eliminate degeneracy when comparing the BCG and ICL to each other and the other cluster components. As detailed in Section \ref{sec:trans_icl_bcg}, authors usually use a surface brightness cut \citep[e.g.,][]{Furnell2021} or a graphical interpretation of the light profile of the BCG+ICL system \citep[e.g.,][]{Kluge2021} to identify the transition region between the BCG and ICL. We opted for the latter method using the same cut of 30 kpc as in \citet{Kluge2021}. We measure The BCG from 5 kpc to 30 kpc to avoid issues of pixel saturation and mergers in the core. We measure The ICL from 30 kpc up to an image specific detection threshold at which the ICL becomes indistinguishable from the background noise and, as such, we are unable to reliably fit an elliptical shape to the distribution. As demonstrated in the radial profiles of each object illustrated in Figures \ref{fig:rogues1}, \ref{fig:rogues2}, and \ref{fig:rogues3}, the radial profile rapidly flattens at $\sim$30 kpc. In the same Figures, we included two 300 kpc $\times$ 300 kpc cutouts of the \emph{HST} WFC3 imaging data for each object. They both cover the exact same field of view except we tuned the left image to highlight the BCG and field galaxies and we tuned the right image to show the low surface brightness features including the ICL. From this, we demonstrate the difference between the two distributions, including our selected cut to differentiate between them. If one were to consider the BCG+ICL system as a whole instead of attempting to differentiate between the BCG and ICL, then, in agreement with the conclusions outlined in Section \ref{sec:con}, they should exclude the inner 30 kpc of their BCG+ICL system when looking for a description of the DM distribution of the cluster in the absence of a better proxy for the DM halo. 

\end{appendix}

\end{document}